\documentstyle[11pt,aaspp4b]{article}
\newcommand{\ts}{\thinspace}
\newcommand{\p}{$^{\prime}$\ts}

\lefthead{Surace et al.}
\righthead{Tip/Tilt Imaging of Warm ULIGS}

\begin{document}

\title{High Resolution Tip/Tilt Near-Infrared Imaging of Warm Ultraluminous Infrared 
Galaxies}

\author{Jason A. Surace}
\affil{Infrared Processing and Analysis Center, MS 100--22, California Institute of 
Technology, Jet Propulsion Laboratory, Pasadena, CA 91125 \\
Electronic mail: jason@ipac.caltech.edu} 
\author{D. B. Sanders}
\affil{University of Hawaii, Institute for Astronomy, 2680 Woodlawn Dr., 
Honolulu, HI, 96822 \\
Electronic mail: sanders@ifa.hawaii.edu}

\centerline{\it To appear in The Astrophysical Journal} 

\begin{abstract}

We present high spatial resolution (FWHM $\approx$ 0.3--0.5\arcsec) H 
(1.6$\mu$m) and K\p (2.1$\mu$m) images of a complete sample (12) of 
ultraluminous infrared galaxies (ULIGs) chosen to have ``warm'' mid-infrared colors 
($f_{25}/f_{60} > 0.2$) known to be characteristic of active galaxies. The extended 
underlying galaxy is detected in each system at H and K\p, as are tidal features and many 
of the star forming knots seen at optical wavelengths. While some of these knots have a 
considerable near-infrared excess, we show that they are likely to have bolometric 
luminosities more similar to that of the extended starbursts with similar optical 
morphology seen in less far-infrared luminous interacting systems. We find that each 
ULIG has increasing contributions at long wavelengths by a very compact source which 
we identify as an active galactic nucleus (AGN). We show that the optical/near-
infrared
colors of these putative nuclei are much more extreme than the most active starburst 
IRAS galaxies, yet are identical to ``far-infrared loud'' quasars which are in turn 
similar to optical quasars combined with large quantities of hot ($\approx$800 K) dust. 
Half of the ULIGs have nuclei with dereddened near-infrared luminosities comparable to 
those of QSOs, while the other half have dereddened luminosities more similar to 
Seyferts, although this may be an effect of patchy extinction and scattering.

\end{abstract}


\noindent Author's note --- due to space considerations in this preprint, the figure
quality has been downgraded. A higher quality preprint can be obtained from
http://spider.ipac.caltech.edu/\~\ jason or by writing the author directly.

\section{Introduction}

One of the most important results from the {\it Infrared Astronomical 
Satellite}\footnote{The Infrared Astronomical Satellite was developed and 
operated 
by the US National Aeronautics and Space Administration (NASA), the 
Netherlands 
Agency for Aerospace Programs (NIVR), and the UK Science and Engineering 
Research 
Council (SERC).} ({\it IRAS}) all-sky survey was the discovery of a 
significant population of galaxies that emit the bulk of their luminosity in the 
far-infrared (e.g. Soifer et al. 1984). Studies 
of the properties of these ``infrared galaxies'' showed systematic trends 
coupled to the total 
far-infrared luminosity; more luminous systems were more likely to appear to 
be 
merger remnants or interacting pairs, and were more likely to possess AGN-like 
emission line features.  A more complete review of the properties of 
luminous infrared galaxies can be found in Sanders \& Mirabel (1996). 
Much recent attention has been focused on ultraluminous infrared galaxies 
(ULIGs), 
objects with infrared luminosities, $L_{\rm ir}$\footnote{$L_{\rm ir} \equiv 
L$(8--1000$\micron)$ 
 is computed using the flux in all four {\it IRAS} bands according to the 
prescription 
given in Perault (1987); see also Sanders \& Mirabel (1996).  Throughout this 
paper we 
use $H_{\rm o}$ = 75 km s$^{-1}$Mpc$^{-1}$, $q_{\rm o}$ = 0.5 (unless 
otherwise noted).}, greater than 
$10^{12}\ L_{\sun}$, which corresponds to the bolometric luminosity 
of QSOs 
\footnote{Based on the bolometric conversion {\it L}$_{\rm bol}$ = 16.5 $\times 
\nu${\it L}$_{\nu}$(B) of Sanders et al.(1989) for PG QSOs. Elvis et al.(1994) 
indicates a value of 11.8 for UVSX QSOs, increasing {\it M}$_{\rm B}$ to -22.5.} ( 
assuming the blue luminosity criterion $M_{\rm B} < -22.1$, adjusting for our 
adopted cosmology
: Schmidt \& Green 1983). 
Multiwavelength 
observations of a complete sample of 10 ULIGs led Sanders et al. (1988a) to 
suggest that these objects might plausibly represent the initial dust-
enshrouded 
stage in the evolution of optically selected QSOs, and that the majority, if 
not all QSOs may begin their lives in such an intense infrared phase. 

An important subset of ULIGs are those objects with ``warm'' mid-infrared colors
($f_{25}/f_{60} > 0.2$).
\footnote{The quantities $f_{12}$, 
$f_{25}$, $f_{60}$, and $f_{100}$ represent the {\it IRAS} flux densities in Jy 
at 12{\ts}\micron, 25{\ts}\micron, 60{\ts}\micron, and 100{\ts}\micron\ 
respectively.}
These warm objects, which 
represent $\sim${\ts}20--25{\ts}\% of the total population of ULIGs discovered by 
{\it IRAS}, appear to 
represent a critical transition stage in the evolution of the larger 
population of ``cool'' ULIGs into optical QSOs.  Studies 
of several small but complete samples of warm ULIGs have shown that many of 
these objects have a point-like optical appearance on the Palomar Sky Survey 
and that they exhibit broad (i.e. Seyfert 1) optical emission lines, characteristics 
that have led them to be referred to as ``infrared QSOs'' (e.g. Low et al. 
1988, Sanders et al. 1988b). This is not surprising since this criterion has been used 
previously to select AGN from the IRAS Point Source Catalog (de Grijp et al. 1985, 
1987). Particularly useful for study has been the 
complete flux-limited sample of 12 warm ULIGs from the survey of Sanders et 
al. 
(1988b). As the nearest and brightest warm ULIGs, these objects are the most 
amenable to studies at other wavelengths. Other selection techniques (e.g. 12$\mu$m 
flux selection; Spignolio \& Malkan 1989) have produced similar, albeit smaller 
samples of ULIGs with AGN spectra.

Surace et al. (1998; Paper I) examined the warm ULIG sample of Sanders (1988b) 
using {\it HST}/WFPC2 and found that nearly all were advanced merger remnants with 
extremely complex optical morphologies in their central few kiloparsecs. These central 
regions were dominated by knots of powerful star formation plus one or two compact 
emission sources whose optical colors, luminosities, and physical sizes appeared similar 
to reddened quasars, and were therefore presumed to be putative active galactic nuclei 
(AGN). Unfortunately, the unknown effects of extinction on small spatial scales resulted 
in considerable uncertainty in the properties of the star-forming knots and putative 
AGN, and optical colors alone could not effectively differentiate between the two 
phenomena.

Extending wavelength coverage to the near-infrared offers several potential 
improvements over previous studies at optical wavelengths.
The most important of these is lowered sensitivity to extinction effects; the extinction at 
2.1$\mu$m is only one-tenth that at 0.5 $\mu$m. In these objects, which are expected 
to have some regions of significant dust obscuration, the near-infrared offers a much 
greater opportunity to detect additional emission regions which may be so heavily 
obscured as to be invisible at optical wavelengths. While some recent studies of 
ULIGs have found that their central regions are opaque even at near-infrared 
wavelengths (e.g. Goldader et al. 1995), these studies have concentrated primarily on 
``cooler'' systems which are likely to be more heavily obscured. The presence of optical 
Seyfert lines in the warm ULIG sample would seem to indicate that extinction is much 
less severe than in the cooler ULIGs.  Also, determination of the near-infrared spectral 
energy distribution (SED) of the optically observed small-scale structure allows a 
disentanglement of the effects of extinction, star-formation, and AGN activity. Color 
information at longer wavelengths is particularly important since this is where normal 
stellar populations and AGN have very different colors and because thermal emission 
from hot 
dust first becomes detectable longward of H-band. 

High resolution observations greatly increase point source sensitivity; in the 
background limited case doubling the spatial resolution is twice as effective as doubling 
the telescope aperture. The increased point source sensitivity  in turn enhances the 
detectability of features such as star-forming knots. The distributed spatial morphology 
of these star-forming knots allows them to be distinguished from AGN. By identifying 
these star-forming regions (some of which may be hidden from optical detection by dust 
extinction) and characterizing their SEDs in the near-infrared, a more accurate 
assessment of the contribution of star-formation to the total ULIG energy budget may be 
obtained.
Unfortunately, the characteristic size-scale of the optical structure is too small 
($\approx$0.2--0.4\arcsec) to be resolved by conventional means even at the best of 
ground-based sites. However, recent advances in adaptive optics (AO) techniques and 
deconvolution, which compensate for local and atmospheric image distortions, allow this 
kind of resolution to be achieved from the ground. We report here the results of such a 
high spatial resolution near-infrared imaging study using a low-order tip/tilt system.

\section{Data}

The data were taken between October, 1995 and May, 1997 with the UH 2.2m telescope 
on Mauna Kea. The observations were made at H (1.6$\mu$m) and K\p (2.1$\mu$m). 
The H filter was chosen since it is the longest wavelength filter which is still relatively 
unaffected by thermal dust emission; dust hot enough to emit significantly at this 
wavelength would be above the dust sublimation temperature. The choice of the 
University of Hawaii K\p filter, which is bluer than both the K-filter and the 2MASS 
K$_S$, was motivated by the lower thermal sky background in the K\p-band 
(Wainscoat \& Cowie 1992). This improves the detectability of faint features as 
star-forming knots. Throughout this paper we exclusively refer to K\p. Comparison to 
work by other authors is made using the conversions (Wainscoat \& Cowie 1992):

\begin{equation}
K-K^{\prime}=0.18 \ (H-K)
\end{equation}

\begin{equation}
(H-K^{\prime})=0.82 \ (H-K)
\end{equation}

A fast tip/tilt f/31 secondary, capable of correcting for image motion at a rate of 10--
100 Hz, was used to stabilize the images against both correlated seeing effects such as 
telescope wind shake as well as some atmospheric seeing (Jim 1995). The detector was 
the QUIRC 1024x1024 camera utilizing a HAWAII HgCdTe array (Hodapp et al. 1996). 
At the f/31 focus, it's 0.06\arcsec pixel$^{-1}$ scale ideally samples the focal plane 
under 
diffraction-limited conditions (0.21\arcsec at 2.1$\mu$m). Because of the camera's 
relatively large field of view ($\approx$61\arcsec), it was possible to dither the 
observations on size-scales larger than the detectable galaxies themselves and 
subtract successive exposures in order to correct for the sky background, thus doubling 
telescope efficiency. The actual pattern used depended on the structure of the galaxy. 
Pixel-pixel response variations were corrected by dividing the sky-subtracted images 
by 
high S/N normalized dome flats. Based on poisson statistics the dome flats are expected to 
be accurate to greater than 0.1\% and the stability of the array generally allows 
flattening with residuals far below 1\%; in the near-infrared one is almost always more 
limited by the bright background sky. The dithered frames were registered using the 
centroids 
of the background stars and galaxies in the frames or in some cases with cross-
correlation using the extended features in the ULIGs themselves. In both cases the 
registration should be accurate to within $\approx$ 0.25 pixels. The images were scaled 
by their exposure times and any constant residual background was subtracted. They were 
then combined using a clipping algorithm in IRAF/IMCOMBINE that rejects high-sigma 
outliers and weights 
the images by their exposure times. Typical total exposure times were 30--120 
minutes per target per filter. In general, any given target was observed using the same 
exposure times. Different exposure times were necessitated by the brightness of the 
objects and the seeing conditions at the time. The QUIRC camera has a very linear 
response over nearly its full range; all exposures were well within the linear regime.

The data were flux calibrated by observing several photometric standards (Elias et al. 
1982, Casali \& Hawarden 1992) throughout the night at the same airmass as the 
targets. In most cases the nights were photometric with a characteristic residual scatter 
of $<$ 0.05 magnitudes. For those where they were not, we 
calibrated the data using either earlier photometric observations at the Palomar 5m 
(taken with a 58x62 InSb array between October 1988 and March 1992) or the 
aperture photometry of Neugebauer et al. (1987) and Sanders (1988a). The K\p{\ts} 
filter was calibrated using the same infrared standards and the above {\it K$-$K\p} 
correction.

The point spread function (PSF) was calibrated in most cases by using actual stars in the 
final combined image. Point sources were identified by eye and the DAOPHOT package in 
IRAF was then used to construct a model PSF for each image. This models consists of a 
theoretical elliptical gaussian core and a ``correction image'' representing the observed 
variations from the theoretical model. In a few cases there were no suitable stars in the 
vicinity of the target. In these cases, a PSF was obtained by interleaving observations of 
the ULIG and nearby bright stars by using the same guide star (and by extension, nearly 
the same degree of image stabilization) for both sets of observations. Although the 
derived PSF is not necessarily identical to that of the ULIG data, there is evidence that 
the atmospheric variability on Mauna Kea is stable on timescales of tens of minutes 
(Ramsay et al. 1992), comparable to the interleave times used. Typical of most AO 
systems, the observed PSF consists of two components: a diffraction-limited core and an 
extended halo approximately 0.5--1\arcsec{\ts} in diameter, equivalent to the 
uncorrected PSF width (Figure 1). This extended halo is due to scattered light in second-
order and higher distortion modes, which cannot be compensated for by a tip/tilt-type 
system (Beckers 1993). Although on a small telescope (such as the UH 2.2m) tip/tilt is 
capable of delivering nearly all the gain in resolving power that can be achieved with 
higher-order systems, this uncompensated extended halo can still contain a significant 
fraction ($\approx$ 30\%) of the total flux, thus complicating photometry. All 
photometry was performed using aperture photometry in circular or polygonal 
apertures, depending on the shape of the observed structure. Aperture corrections were 
derived from the observed PSF.

\begin{figure}[tbhp]
\epsscale{0.8}
\plotone{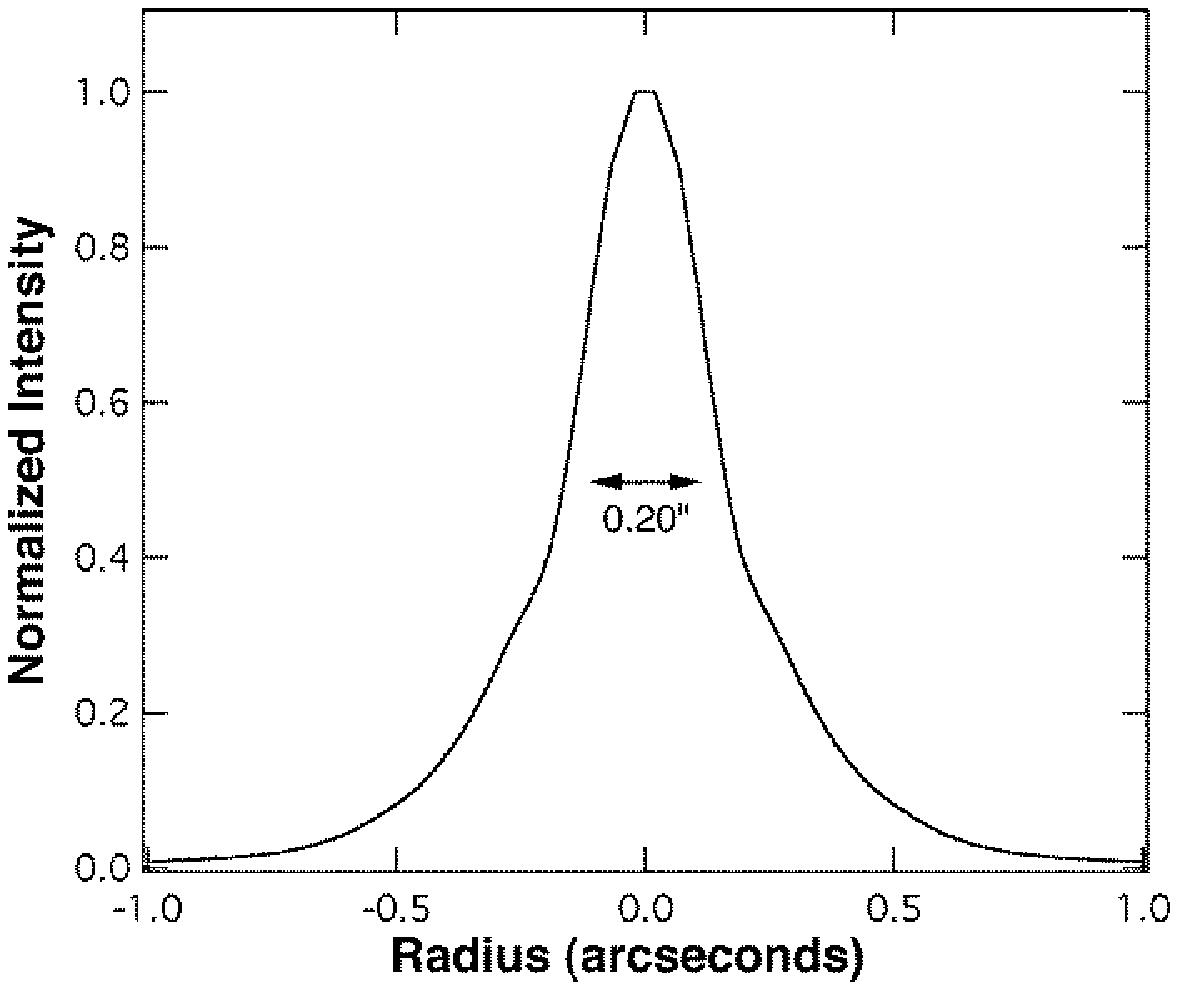}
\caption{Typical radial profile of the point spread function at K\p achievable
with fast
tip/tilt guiding. A FWHM of $\lambda$/D at K\p has been denoted by the scale-bar
. Note
that the central core is nearly diffraction-limited in width, but that a considerable
amount of power remains in the wings due to higher-order scattering modes.}
\end{figure}

In several cases, deconvolution was applied. This was particularly necessary in those 
cases where structure was seen on very small spatial scales ($< 0.5$\arcsec{\ts}). 
Deconvolution is able to increase the contrast and detectability of small-scale structure 
and improve the photometry by eliminating confusion resulting from the extended 
scattered halo. We chose to employ the Richardson-Lucy algorithm (R-L) implemented 
in IRAF. The algorithm was allowed to iterate 25 times. As a check on the validity of the 
deconvolved structure, we compared the deconvolved images with the {\it HST} optical 
images, which are of much higher resolution. In general they have similar structure, 
and will be discussed more in \S 3.1. In all cases photometry was performed on both 
the raw and deconvolved data.

In the case of one galaxy, IRAS{\ts}07598+6508, it has not been possible to obtain 
complete, high quality data at both H \& K\p. Therefore, while images and radial profiles 
at J 
and K-band are presented, it will be excluded from further discussion. It has a Seyfert 1 
optical nucleus, leading some authors to describe it as an ``infrared quasar'' (Low et al. 
1988), as well as considerable tidal debris in the surrounding host galaxy (Boyce et al. 
1996). Since it is widely regarded as a QSO, it will therefore be considered similar to 
the three ``infrared-loud'' QSOs newly observed at H \& K\p and discussed in \S 3.3.4. 

\section{Analysis}

\subsection{Near-Infrared Morphology}

{\it HST}/WFPC2 imaging indicates that the optical emission of these warm ULIGs is 
characterized by a dominant nucleus which is the most compact (generally unresolved by 
{\it HST}) and luminous source at {\it B} and which has been identified as a putative 
AGN, as well as by compact star-forming regions (with the exception of 3c273) found 
predominantly in the inner few kpc (Paper I). In most cases these star-forming regions 
seem to have ages based on their optical colors of less than $\approx$ 5$\times$ 
10$^8$ years. The near-infrared, however, traces primarily the late-type stellar 
population at H-band with increasing sensitivity to emission from hot dust at K\p-band 
(Aaronson 1977). 
Figure 2 shows the large scale structure observed at H-band for the complete sample of 
warm ULIGs. The near-infrared morphology generally appears to be much simpler than 
that seen optically.
In every object, the extended emission of the surrounding galaxy is detected and in 
several 
cases we can even trace the tidal structure (e.g. Mrk 1014, IRAS{\ts}05189$-$2524, 
IRAS{\ts}08572+3915, Mrk 231, Pks{\ts}1345+12, and Mrk 463). Three of the 
objects (IRAS{\ts}08572+3915, Pks{\ts}1345+12, and Mrk 463) show clear double 
galaxy nuclei, a result known from previous optical and near-infrared imaging (Sanders 
1988a; Carico et al. 1990).
Additionally, we can detect the spiral arms and star forming knots of the host galaxy of I 
Zw 1, the star-forming knots in the outer regions of Mrk 463 and 
IRAS{\ts}15206+3342, and the extended host of IRAS{\ts}12071$-$0444. However, 
only in a few cases (IRAS 15206+3342, Mrk 231, and Mrk 463w) is there any 
indication of complex near-infrared structure in the immediate vicinity of the nuclei, 
and even then this structure is comparatively faint compared to the nuclei themselves 
(Figure 3). This smoothness is in part due to the lower resolution of the near-infrared 
images. However, the typical size scale of the optical knots is 150 pc, with a knot 
separation of 300--500 pc. In this case we would expect many of the knots to be 
separated. Typical {\it B$-$K\p} colors for a starburst lie in the range 0--4, with {\it 
B$-$
K\p}$\approx$ 2.5 for a starburst 10$^8$--10$^9$ years in age. Given that the 
starburst knots detected in Paper I were typically {\it m}$_{\rm B}$=22--24, then 
most of the knots should be {\it m}$_{\rm K^{\prime}} \approx$19.5--21.5. The 
detection limit for point sources is typically {\it m}$_{\rm K^{\prime}}$=22--23; 
most of the optical knots should therefore be detectable. Unfortunately, matters are 
complicated by the more limited spatial resolution in the near-infrared, which limits 
the dynamic range in the vicinity of very strong point sources, and also by confusion due 
to the much stronger emission in the near-infrared from the underlying host galaxy. The 
former may not actually be very significant, given the nearly unresolved nature of the 
K-band peaks (see below).

\begin{figure}[p]
\vspace{13.0cm}
\caption{The large-scale structure observed at H. The data have been adaptively
smoothed, a process which applies greater smoothing to regions with low {signal-
to-
noise} in order to increase contrast. The data are displayed such that the gray-
scale part
of the image illustrates low surface brightness features, while the white logarithmic
contours show the high surface brightness structure. Tick marks are 0.5 \arcsec
apart,
with major ticks every 4 \arcsec. Northeast is at top left. {\bf This figure is included as a separate JPEG file.}}
\end{figure}

\begin{figure}[p]
\vspace{13.0cm}
\caption{Selected K\p (Mrk 231 and IRAS{\ts}15206+3342) and H-band (Mrk
463) data deconvolved with the Richardson-Lucy algorithm to show high spatial
frequency structure near the nuclei. The indicated knots are spatially coincident
with the
optical knots of Surace et al. (1998), and have been labeled with their optical
names.
The one new infrared knot is labeled ``irA''. Northeast is at top left. {\bf This
figure is included as a separate JPEG file.}}
\end{figure}

Despite the more limited spatial resolution in the near-infrared, the star-forming 
knots seen optically are clearly detected in Mrk 231, Mrk 463, and 
IRAS{\ts}15206+3342. Figure 3 displays these high spatial-resolution, high surface 
brightness features, which have been enhanced by Richardson-Lucy deconvolution. These 
bright, compact emission regions are spatially coincident with the optical knots, and it 
seems likely that the near-infrared emission also arises from these knots.
The many knots detected optically in IRAS{\ts}08572+3915w were not detected in the 
near-infrared, despite the near-infrared observations being deep enough to see {well-
defined} tidal tails. It seems likely that this is due to a combination of heavy obscuration 
towards the infrared nucleus and the blueness of the optical knots. 
Even at optical wavelengths, this region's appearance at B and I is so different as to 
preclude easy 
registration of features. At near-infrared wavelengths it's appearance is that of a single 
source unresolved at K\p; the optical knots are too faint to avoid confusion with the 
bright nucleus. 

Perhaps most surprising is that almost no {\it new} knots have been discovered in the 
warm ULIGs. It was argued in Paper I that the optically observed star-forming knots 
were insufficiently luminous to be a significant contributor to the high bolometric 
luminosity in these ULIGs. However, one possibility is that a significantly larger 
population of star-forming knots exists, yet is embedded deeply enough so as to be 
undetectable in the WFPC2 data (limiting magnitude {\it m}$_{\rm B} <$ 27). 
Assuming that most of the knots have a similar intrinsic luminosity, any knot 
extinguished by more than {\it A}$_{\rm V}$= 4 magnitudes would have been 
undetectable by {\it 
HST}. However, the extinction at K\p is only 1/12 that at B; knots which were 
extinguished just enough to be undetectable at B would be extinguished by only 0.3 
magnitudes at K\p. Given the K\p detection limit and an expected K\p magnitude of 20 for 
the brightest knots, similar knots would be seen even if they were obscured by{\it 
A}$_{\rm V}$ = 25 magnitudes. This leads to the conclusion that either there is no 
population of 
moderately embedded star-forming knots in these galaxies or that any moderately 
embedded clusters are intrinsically less luminous than those seen optically. If any more 
knots exist like those seen optically, then they have line-of-sight extinctions greater 
than {\it A}$_{\rm V}$ = 25 magnitudes. The failure to find any large populations of 
spatially 
distributed, moderately embedded knots lends credence to the idea that the dust-clearing 
time in the vicinity of the knots is short compared to the current starburst lifetime. It 
additionally supports the finding of Paper I that the knots do not contribute greatly to the 
bolometric luminosity, since there are relatively few of them, at least to an optical 
depth of {\it A}$_{\rm V}$= 25 magnitudes.

An examination of the radial profiles in both bands (Figure 4) is particularly 
instructive. In nearly all cases, the ULIGs are extended on large scales (1 \arcsec 
$\approx$ 1 kpc) at H. The underlying old stellar population of the merger remnant 
thus seems to contribute strongly at H. The only notable exceptions are three systems 
already known to be dominated even at optical wavelengths by an unresolved central 
source: Mrk 1014, IRAS 07598+6508, and 3c273. All three are QSOs (Schmidt \& 
Green 1983), and even in these cases, all three are known to possess sizable host 
galaxies (McLeod \& Rieke 1994b, Boyce et al. 1996). The high luminosity of their 
central peaks is sufficient to effectively mask the low surface brightness hosts at the 
depth and radii illustrated in Figure 4. In contrast to H, at K\p many of the ULIGs are 
dominated by a single, essentially unresolved point source. This difference in 
compactness between H and K\p is particularly evident in IRAS{\ts}05189$-$2524, 
IRAS{\ts}08572+3915w, IRAS{\ts}12071$-$0444, Pks 1345+12w, and Mrk 463e. 
The three QSOs (Mrk 1014, IRAS{\ts}07598+6508, and 3c273) show this effect to a 
lesser degree as they are already dominated by a central point source at H and K\p. Of the 
remaining galaxies, I Zw 1, also a QSO (Schmidt \& Green 1983), is superimposed on an 
extremely large and luminous host, and hence has a considerable host contribution even 
at small radii. The Seyfert 1 galaxy Mrk 231 is similarly known to possess a central 
``monster'', yet it is also superimposed on an extremely luminous host. This leaves 
IRAS{\ts}15206+3342 as the only warm ULIG which is either not already known to be 
dominated by a central point-like source, or does not show increasing compactness at 
long wavelengths indicating that the  majority of the bolometric luminosity is probably 
also compact. However, a single peak (the putative nucleus in Paper I) does have 
noticeably redder near-infrared colors than the other dozen or so detected knots. This 
may suggest that while this ``nucleus'' does not yet dominate as strongly at 2.1$\mu$m 
as in the other ULIGs, it may begin to do so at slightly longer wavelengths.

{
\begin{figure}[p]
\plotone{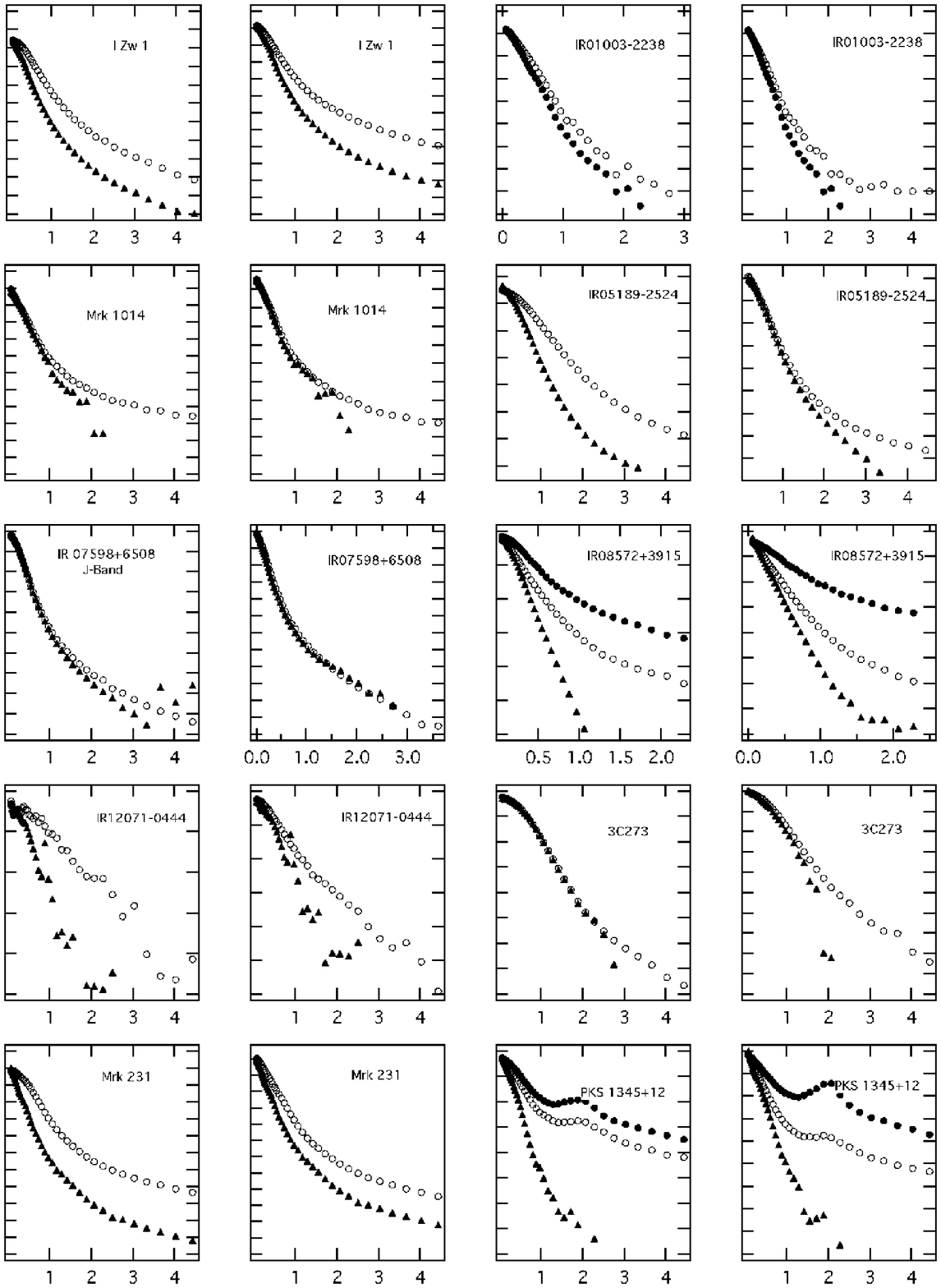}
\caption{Radial profiles at H (left) and K\p (right) of the warm ULIGs (eastern
nuclei=filled
circles; western nuclei=open circles) and the associated point spread functions
(triangles), plotted as normalized surface intensity (in magnitudes) versus radius
 (in
arcseconds).
The galaxies are all noticeably extended at H, while they are essentially unresolved
at K.
Note that in terms of surface brightness, FWHM intensity occurs at just 0.3
mag/\arcsec{\ts} below the peak.}
\end{figure}
\addtocounter{figure}{-1}
\begin{figure}[t]
\plotone{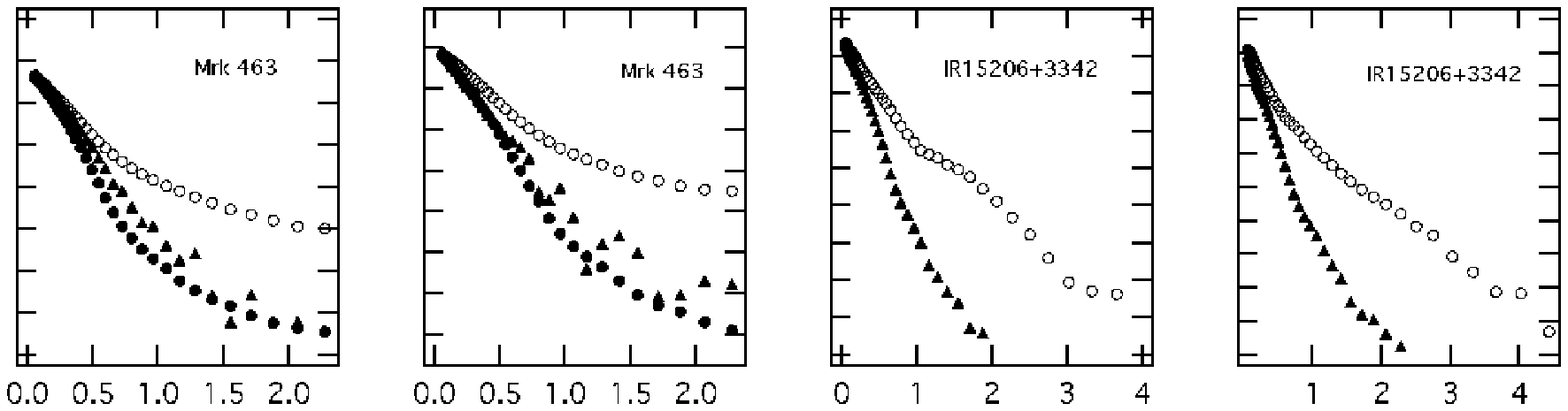}
\captionnoentry{(continued)}
\end{figure}
}

The physical FWHM of the PSF at K\p ranges from 300 pc in Mrk 231 to nearly 1.5 kpc 
in Mrk 1014, with a median of 600 pc. In most cases the emitting region at K\p 
(which appears to be spatially coincident with the optical nucleus) is at least this small. 
This is not surprising since the emission region is expected to be very small if 
the ULIGs harbor a dust-enshrouded QSO. Similarly, the ultracompact starbursts 
postulated by Condon et al. (1991) are also smaller than we can resolve in the near-
infrared, and the putative nuclei identified in Paper I also have size scales much smaller 
than the near-infrared resolution limit. This result implies that the majority of the K\p 
-band flux in the warm ULIGs is emitted from a region spatially distinct from the H-
band and optical emission. This 
increasing compactness at long wavelengths indicates a strong infrared excess on small 
spatial scales, which we show in \S 3.3 is most easily explained by the presence of hot 
dust.

\subsection{Near-Infrared Luminosities}

The photometry for the total integrated galaxy light and the putative nuclei are presented 
in Table 1.
The integrated galaxy magnitude represents all of the light from both the galaxy and the 
putative nuclei. It was measured using a 30\arcsec{\ts} diameter aperture, 
corresponding to 30--100 kpc; in all cases this was sufficiently large that the observed 
galaxy emission at the edge of the aperture was no longer visible against the background 
noise. The H-band luminosity of the putative nucleus was determined in two ways. First, 
detailed {two-dimensional} PSFs were fit to the galaxies; the resulting fitted PSFs were 
integrated in order to estimate the nuclear flux. This technique has several 
shortcomings; in particular, the fit is complicated by an ignorance of the underlying 
intrinsic galaxy profile. As an estimate, the fit was made by scaling the PSF until the 
galaxy profile reversed itself, i.e. a ``hole'' was dug in the nuclear regions. The other 
technique was simply to measure the total flux inside a fixed 2.5 kpc diameter aperture. 
Although this technique ignores the underlying nuclear galaxy emission, it is fairly 
successful in situations where the central regions are dominated by a bright point 
source. In these cases, the surface brightness of the underlying galaxy is so low inside 
the fixed aperture compared to that of the nucleus such that it represents only a small 
fraction of the total flux. It was found that in nearly all cases, the two techniques agreed 
with each other at approximately the 10\% level. Comparison of the photometry with 
that of previous studies (Sanders et al. 1988, Mcleod et al. 1994a, 1994b) shows that 
it is the same to within 15\% .


\begin{deluxetable}{lrrrrrrr}
\footnotesize
\tablenum{1}
\tablewidth{6.0truein}
\tablecaption{ULIG Total and Nuclear Photometry}
\tablehead{
\colhead{Name} &
\colhead{{\it m}$_{\rm H,total}$}&
\colhead{\it m$_{\rm H,nuclear}$}&
\colhead{${L_{\rm nuclear}}\over{L_{\rm total}}$}&
\colhead{\it M$_{\rm H,host}$}&
\colhead{\it m$_{\rm K^{\prime},total}$}&
\colhead{\it m$_{\rm K^{\prime},nuclear}$}&
\colhead{${L_{\rm nuclear}}\over{L_{\rm total}}$}
}
\startdata
I Zw 1$^{\dagger}$& 11.19 & 11.69 & 0.63 & -24.5 & 9.94 & 10.52 & 0.53 \nl
IRAS 01003-2238 & 15.03 & 15.80 & 0.49 & -22.7 & 14.30 & 15.33 & 0.39 \nl
Mrk 1014$^{\dagger}$&12.84& 13.53 & 0.53 & -25.6 &11.67& 12.28 & 0.57	\nl
IRAS 05189-2524        &11.21 &	11.93 & 0.52 & -24.2 &10.23 &10.44 & 0.82	\nl
IRAS 07598+6508 & 13.01 & \nodata & \nodata & \nodata & 10.45 & \nodata & \nodata \nl
IRAS 08572+3915W      &13.65 &	15.17& 0.25 & -22.9 & 12.73 &13.75 & 0.39 \nl
IRAS 12071-0444	&	14.28 & 15.35 & 0.37 & -23.9 &13.42 & 13.91 & 0.64 \nl
3c273$^{\dagger}$	& 10.88  & 11.04 & 0.86 & -26.1 & 9.70 & 9.85 & 0.87 \nl
Mrk 231	               &9.91  &	10.13& 0.82 & -24.4 & 8.87 & 8.98 & 0.90  \nl
Pks 1345+12W           &12.99 &	15.36& 0.11 & -25.5 & 12.62 & 14.21 & 0.23\nl
Mrk 463E               &11.19 &	12.05 & 0.45 & -24.8 & 10.20 & 10.50 & 0.76 	\nl
IRAS 15206+3342        &14.14 & 15.40 & 0.29 & -24.1 & 13.47 & 14.58 & 0.36 \nl
\tablecomments{$^{\dagger}$ calibrated using photometry of Neugebauer et al. 1987. IRAS 07598+6508 of poor quality .
Uncertainty in total magnitude is 0.05 magnitudes, except for those marked with a $\dagger$, where it is 0.1 magnitudes. Uncertainty in the nuclear luminosity is 0.1 magnitudes.}
\enddata
\end{deluxetable}

The emission at H-band is dominated by the old stellar population. The presence of double 
nuclei and well-formed tidal tails in many of these systems indicates that the merger 
process is 
not yet complete, and simulations indicate that they are unlikely to be more than 1--2 
Gyr (Barnes 1992, Mihos \& Hernquist 1994) in age. Therefore the H-band emission is 
indicative of the current state of the stars 
which predated the merger itself, and the integrated H-band luminosity can be used to 
characterize the progenitors' masses by assuming a common {\it M/L}$_{\rm H}$ ratio 
and comparing them to a typical spiral galaxy. Characterization of the mass of the 
underlying ``host'' in the warm ULIGs is interesting for several reasons ( here we define 
the word ``host'' to mean the smooth component of the galaxy luminosity profile which is 
unlikely to be a result of clustered star-formation or non-stellar nuclear activity). In 
particular, it will allow us to determine if the ULIGs represent the collision of typical 
gas-rich spirals (Sanders et al. 1988a), or whether they are peculiar. For example, the ULIGs might be scaled-
up versions of their less luminous counterparts, and simply represent the chance 
collision of unusually massive galaxies, in which case we would expect to find an 
unusually luminous host galaxy at H. Similarly, it may be that in order to drive enough 
material efficiently into the merger nucleus to trigger ULIG activity, the colliding galaxies must 
have very large masses. 
This is also important in testing any possible evolutionary scenario linking ULIGs and 
optically-selected QSOs; if QSOs are the evolved remnants of ULIGs, then they must have 
similar total old stellar masses and hence H-band luminosities.

Table 1 presents the derived photometric quantities: the nuclear fraction of emission at 
H and K\p, and the H-band luminosity. McLeod \& Rieke (1994a) estimate that for an 
{\it L$^*$} galaxy {\it M}$_{\rm H}$=$-$23.9; adjusting to {\it H}$_{\rm 0}$=75 
kms$^{-1}$, this is {\it M}$_{\rm H}$=$-$24.04 . Using {\it M}$_{\rm V}$=$-
$21.14 for 
an {\it L$^*$} galaxy with {\it H}$_{\rm 0}$=75 kms$^{-1}$ and an estimate of ({\it 
V-H})=2.7 for an aged stellar population with a Salpeter IMF (BC 93), we similarly 
derive M$_H$=$-$23.84. The range in host galaxy luminosities is $\sim$0.4 {\it 
L$^*$} (IRAS{\ts}01003-2238) to nearly 7 {\it L$^*$} (3c273), with a mean of 2.4 
{\it L}$^*$. Six of the ULIGs have hosts in the range 1--2 {\it L$^*$}, while three (2 
QSOs and a powerful radio galaxy) are in the range 5--7 {\it L}$^*$. This seems to be 
consistent with the idea that most of the ULIGs are the result of mergers between typical 
(i.e. {\it L$^*$}) or perhaps slightly more luminous than normal galaxies. Previous 
studies have derived similar results: $\sim$2.5 {\it L}$^*_{\rm B}$ (Armus et al. 
1990, Jensen et al. 1997) and 2.5 {\it L}${^*}_{\rm K^{\prime}}$ (Jensen et al. 
1997) for the BGS ULIGs and ``Arp 220-like'' objects, and $\sim$3 {\it L}$^*_{\rm 
r}$ (ranging from $\sim$ 1--24 {\it L}$^*_{\rm r}$) for the ULIG 2 Jy sample 
(Murphy et al. 1996). Finally, the range and distribution in {\it M}$_{\rm H}$ for the 
``warm'' ULIGs is very similar to that of the host galaxies of PG QSOs (McLeod \& Rieke 
1994a,b), which typically range from 0.5--7 {\it L}$^*_{\rm H}$, with a mean of 
1.5 
{\it L}$^*_{\rm H}$. Additionally, McLeod \& Rieke (1994a,b) note that the host 
galaxies of luminous quasars are more similar to 2 {\it L}$^*_{\rm H}$, and are more 
likely to have tidal features indicative of past galaxy-galaxy interactions.

IRAS 08572+3915 is surprisingly underluminous ($\approx$0.5 {\it L}$^*$) at H 
considering that it's morphology is clearly that of two merging spiral galaxies (Surace 
et al. 1998). A possible explanation for this is the extremely high optical depth 
throughout the western galaxy component. Very red near-infrared colors (0.7 $< H-
K^{\prime} <$ 1.4) within 2 kpc of the even redder nucleus ({\it H-K}$^{\prime} > 
$2) indicate that dust obscuration may be very spatially widespread. Assuming a typical 
old stellar population color of ({\it H$-$K}$^{\prime}$)=0.16, then this would imply 
{\it A}$_{\rm H} \approx$ 2--3. Dereddening this region by 2 magnitudes at H would 
increase the total galaxy brightness by 1.2 magnitudes, making the estimated luminosity 
of the underlying old stellar population much more similar to the range found for the 
other ULIGs at $\approx$ 1.2 {\it L}$^*$. We note that the derived luminosity for 
IRAS{\ts}08572+3915 in K\p -band is nearly 0.7 {\it L}$^*$, further indicating that 
substantial extinction may obscure the old stars.

This illustrates an important point: the luminosities given in Table 1 are {\it lower 
limits} to the true H-band luminosity. As illustrated by IRAS 08572+3915 above, 
there may still be considerable extinction at H due to spatially extended dust. This may 
affect all the ULIGs which have large, spatially extended regions that are redder than 
expected for late-type stars: notably, IRAS{\ts}05189$-$2524, 
IRAS{\ts}08572+3915, 
IRAS{\ts}12071-0444, and Mrk{\ts}463. IRAS{\ts}08572+3915 is the most extreme 
example. 
There may also be very low surface brightness, very extended emission on physical 
scales greater than 50 kpc (e.g. in tidal tails) which we fail to detect in the near-
infrared. This can be evaluated by comparing the spatial extent of the near-infrared data 
to previous optical data ( Sanders et al. 1988a, Surace et al. 1998). In the cases of 
IRAS{\ts}05189$-$2524, IRAS{\ts}{12071$-$0444}, Mrk{\ts}231, and 
IRAS{\ts}15206+3342 there is extended structure seen optically which cannot be 
detected in the near-infrared data. In the remainder of the galaxies, all of the optically 
detected extended structure is detected in the near-infrared, and hence this is probably 
unlikely to affect them.
Finally, the techniques used to determine the nuclear emission component will always 
result in over-subtracting from the underlying galaxy. This effect is most pronounced 
when the contrast between the nuclear component and underlying galaxy core is 
minimal, for example as in IRAS{\ts}05189$-$2524, IRAS{\ts}08572+3915, 
IRAS{\ts}12071$-$0444, and Pks{\ts}1345+12. 

It is interesting to note that none of these sources of error is likely to affect 
IRAS{\ts}01003$-$2238. This object is very blue and pointlike, does not have a high 
surface brightness background host, and has only a few knots of star formation visible 
optically, with no extended tidal features. The presence of strong {Wolf-Rayet} emission 
possibly associated with the optical star-forming knots would seem to preclude this 
being a merger so advanced as to have no visible tidal remnants. It is under-luminous at 
H ($\approx$ 0.3--0.5 {\it L}$^*$), and this, combined with the above, would seem to 
indicate that unlike the other ULIGs it is either not a merger system or it is a merger 
involving under-luminous ( and probably low-mass) galaxies. 

\subsection{Optical/Near-Infrared Colors}

It is possible to examine the spectral energy distribution in the optical/near-infrared 
as characterized by the ({\it B$-$I}), ({\it I$-$H}), and ({\it H$-$K\p}) colors by 
combining 
the new near-infrared data with the optical data of Paper I. This makes possible the 
disentanglement of the effects of extinction and dust emission. As has been noted 
previously, 2-color diagrams suffer a degeneracy in that many different combinations of 
extinction and thermal dust emission can produce similar results thus making it 
difficult 
to determine uniquely the underlying emission mechanism. This additionally complicates 
questions such as the {age-dating} of knots based on colors alone, since even modest 
amounts of reddening can result in large variations in age estimates. 

\subsubsection{Modeled Colors}

The stellar colors are based on the spectral synthesis models of Bruzual \& Charlot 
(1993; the updated version used is hereafter called BC95). The particular model we 
have chosen is an instantaneous starburst with 
a Salpeter IMF and lower and upper mass cutoffs of 0.1 and 125 M$_{\sun}$. Of 
particular importance is that stellar ({\it H$-$K$^{\prime}$}) colors reach a 
maximum 
of roughly 0.16, which corresponds to an extremely old age ($> 10^9$ yrs) for a 
stellar ensemble. This is very similar to the observed colors of typical galaxies 
(Aaronson 1977).

The QSO colors are based on a model of a typical QSO optical/near-infrared SED. The 
continuum emission is modeled with a two-part broken power law derived from 
Neugebauer et al. (1987) for the Bright Quasar Survey (BQS):

\begin{equation}
\begin{array}{cr}
F_{\lambda} \propto {\lambda}^{-(2+{\alpha})} & \ {\rm where}\ \alpha=-1.4 ; 
{\lambda} > 1{\mu}{\rm m} \\
 & \alpha=-0.2; {\lambda} < 1{\mu}{\rm m}\\
\end{array}
\end{equation}

Broad line emission based on the line widths, equivalent widths, and line strengths 
determined by Baldwin (1975), Davidson \& Netzer (1979), Wills et al. (1985) and 
Osterbrock (1989) was also modeled. 
The derived QSO colors as a function of redshift are given in Table 2. The colors derived 
from this model are in good agreement with the observational results of the Elvis 
(1994: UVSX) sample, whose average colors are ({\it B$-$I})=1.1$\pm$0.3, ({\it 
I$-$
H})=1.3$\pm$0.3, and ({\it H$-$K\p})=0.78$\pm$0.2 at a median redshift of 0.15. 
The 
presence of strong line emission is the cause of most of the dependency of the color on 
redshift. In particular, H$\alpha$, H$\beta$, and MgII are shifted through the B and I 
filters in the redshift range $0<z<0.16$. The considerable scatter in QSO colors is 
likely to be intrinsic. Neugebauer (1987) noted a considerable scatter in the continuum 
power-law indices; these alone can easily account for the observed color variances.



\begin{deluxetable}{lrrr}
\footnotesize
\tablenum{2}
\tablewidth{2.5truein}
\tablecaption{Derived QSO optical/near-infrared Colors}
\tablehead{
\colhead{{\it z}} &
\colhead{{\it (B-I)}}&
\colhead{{\it (I-H)}}&
\colhead{{\it (H-K$^{\prime}$)}}
}

\startdata
0.00&	0.54&	1.75&	0.78 \nl
0.05&	0.61&	1.68&	0.78 \nl
0.10&	0.75&	1.50&	0.78 \nl
0.15&	0.93&	1.20&	0.78 \nl
\enddata
\end{deluxetable}

The effects of line-of-sight extinction, i.e. a simple foreground dust screen, were 
derived from Rieke \& Lebofsky (1985). The reddening 
law was linearly interpolated to the rest frame wavelength of the galaxy in order to 
account for redshift. 

The spectrum of hot dust with a characteristic temperature, {\it T}, and a $\lambda^{-
2}$ emissivity law was modeled according to the prescription given by Aaronson 
(1977):

\begin{equation}
F_{\lambda} \propto \lambda^{-2} B_{\lambda} \propto  \lambda^{-7} (E^{{\rm 
hc}\over{{\lambda}{\rm k}T}}-1)^{-1}
\end{equation}

Modeled dust temperatures were in the range 600--1000 K. The effects of cooler 
thermal dust emission at long wavelengths on color are degenerate since the dust 
emission is so red that it only increases the K\p -band flux, and hence the ({\it H$-$
K$^{\prime}$}) color. Only at temperatures above 600 K can dust emission contribute 
appreciably at wavelengths shorter than 2 $\mu$m. 1000 K is near the realistic ceiling 
for dust temperatures; at temperatures above $\sim$1500 K, the dust evaporates 
(Spitzer 1978). In any case, such a hot dust model does not fit the data well as even 
1000 K is clearly hotter than needed.

Free-free emission was modeled according to (Spitzer 1978)

\begin{equation}
F_{\lambda} \propto {\lambda}^{-2} e^{-{{\rm hc}\over{{\lambda}{\rm k}T}}}
\end{equation}

\noindent assuming a 20,000 K electron temperature. The $\lambda^{-2}$ term 
dominates this expression; the slowly varying Gaunt factor is regarded as constant. 
Similarly, at longer wavelengths the exponential term is near unity, and only the short 
wavelength colors are affected appreciably by it. However, these same short wavelength 
colors will be shown to be poor at discriminating the underlying emission mechanism.

Finally, the case of mixed stars and dust was considered. The spatial extent of the knots is 
such that they must be very large associations of many stars. It is both possible and 
likely that there are local, unresolved regions of high extinction inside the knots. An 
example of this might be a small region of star-formation containing several embedded 
protostars. The result of the presence of such unresolved regions inside the knots is to 
increase the 
thermal near-infrared component of the SED relative to that of a naked starburst, in a 
manner similar to that of the hot dust modeled above. Note that this is different from the 
effects of uniform line-of-sight extinction as modeled above. 

We have modeled the effect of mixed stars and dust by assuming an underlying power-
law distribution for the unreddened luminosity {\it L}$_{A_{\rm V}}$ obscured by a 
given extinction {\it A}$_{\rm V}$:

\begin{equation}
L_{A_{\rm V}} = L_{A=0} (A_{\rm V}+1)^{\alpha}
\end{equation}

%

\noindent This is an ``onion-skin'' model; it peels away each optical depth and computes 
the total luminosity in each filter from all emission at that optical depth. The colors 
were derived by computing a numerical model of the observed luminosity at each optical 
depth, and then summing over all extinctions from {\it A}$_{\rm V}$=0 to {\it 
A}$_{\rm V}$={\it A}$_{\rm max}$. The resulting colors are displayed in Figure 5.
The $\alpha$=0 case corresponds to dust mixed uniformly with the emitting source, and 
is similar to that of Aaronson (1977) and Carico et al. (1990). The ({\it B$-$I}) 
colors 
rapidly converge on a fixed value determined by $\alpha$, since the effective optical 
depth rises very rapidly at short wavelengths and hence stars at {\it A}$_{\rm V} >$10 
magnitudes
cannot contribute more than a tiny fraction to the observed short wavelength emission. 
At {\it A}$_{\rm V} >$ 50 magnitudes this becomes true even at K\p, and thus each 
model converges 
at high optical depth on a color dictated by ${\alpha}$. This behaviour can be most easily 
described by the statement that the colors are fixed by the total stellar luminosity at 
optical depths less than one for any given wavelength, i.e. only the relatively unobscured 
stars in the outer regions of the dust/star mixture contribute noticeably to what is 
observed. This is similar to the result of Witt et al. (1992) that any extended stellar 
distribution mixed with a scattering and absorbing interstellar medium reaches a 
maximum reddening value at sufficiently high optical depths. While models with large 
amounts 
of stars at high optical depth (e.g. ${\alpha}$=2) have fairly large ({\it H$-$K\p}) 
colors, they are accompanied by a similar increase in ({\it B$-$I}) and ({\it I$-$H}) 
which 
is not entirely dissimilar to the effects of reddening. In the reddening-orthogonal color 
basis discussed below, nearly all four models lie in essentially the same color location, 
and hence only the $\alpha$=0 case will be illustrated. We have assumed that the 
luminosity source has the same colors as an intermediate age starburst ({\it B$-$I, I$-
$H, H$-$K}\p = 0.62, 0.97, 0.2), and that the dust 
obeys the same reddening law discussed above. The former assumption is almost 
certainly unrealistic, since it ignores the thermal dust emission associated with 
protostars found inside star-forming regions. However, this will be discussed using an 
additional observational model presented in \S 3.4. 

\begin{figure}[p]
\plotone{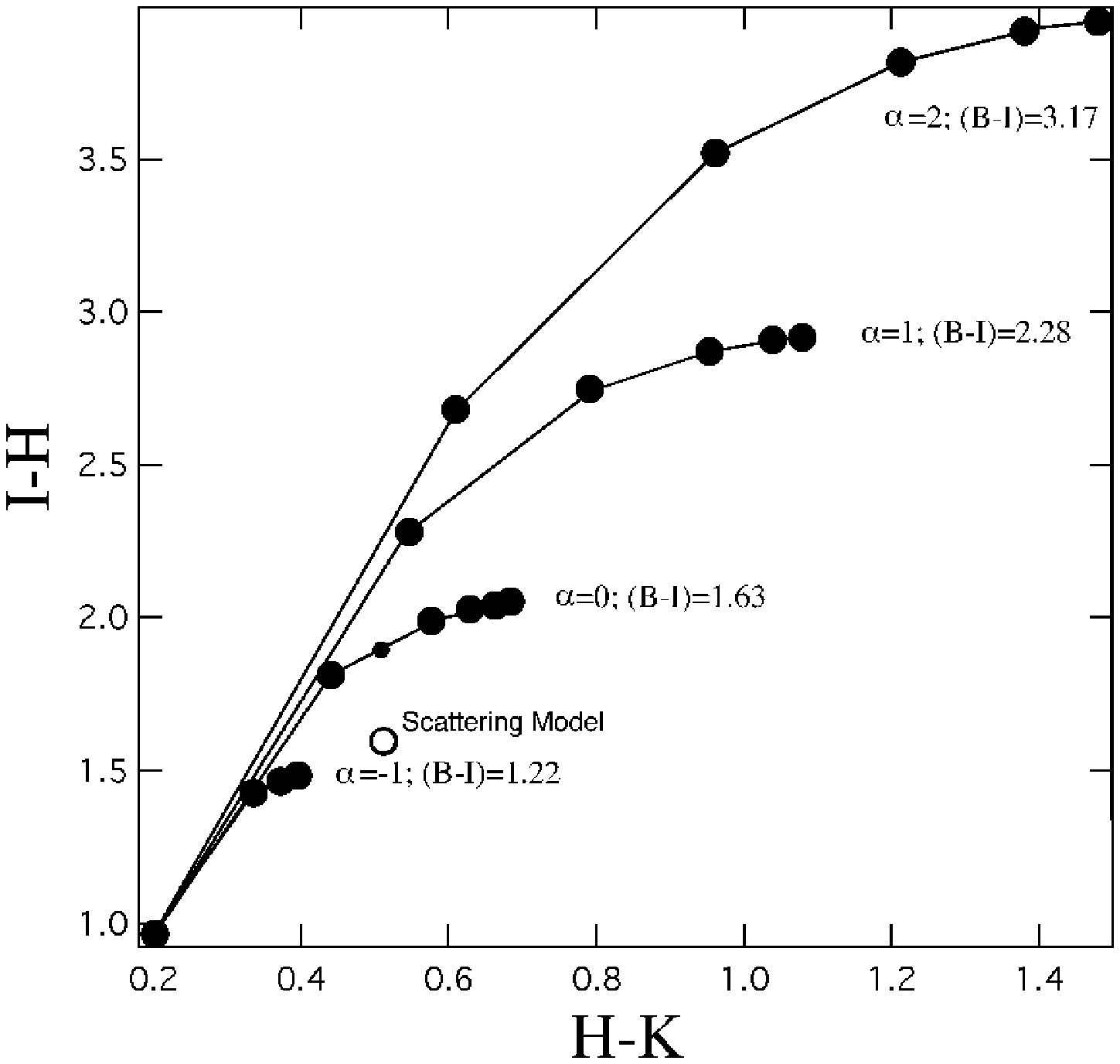}
\caption{({\it I$-$H}) and ({\it H$-$K\p}) colors for models of mixed stars and
local
dust extinction, with the luminosity at each optical depth characterized by a power-law
of index ${\alpha}$. The marked intervals are the maximum optical depth in units
of
A$_{\rm V}$ = 10 magnitudes. The ({\it B$-$I}) colors for a given model become fixed
rapidly since only
stars at low optical depth contribute significantly to the short wavelength emission.
This
is true even at K\p-band at high optical depth, hence the colors converge on a model-dependent
value. The open circle indicates colors for mixed star and dust models
including scattering with {\it A}$_{\rm V}$ = 16 magnitudes, although ({\it B$-$
I})
is lower (1.46 for evenly mixed stars and dust, and 0.77 for centrally concentratated
stars and dust). The same point in the non-scattering models is marked with a small
filled circle.} 
\end{figure}

The treatment described above ignores the effects of scattering, which may be quite 
important for an extended distribution of stars and dust. Witt et al. (1992) present 
models for various spherical distributions of dust and stars, with a full treatment of 
scattering. Of particular interest here are those models for ``dusty'' galaxies in which 
dust and stars are uniformly mixed. This corresponds to the model generated above, and 
is the most likely to represent the situation within the starburst knots (other models 
which represent dusty starburst cores or obscured AGN are more similar to the galaxies 
taken as a whole, rather than as small-scale individual knots). The addition of scattering 
leads to an increase in the fraction of emergent light, particularly at short wavelengths. 
The evolution of optical/near-infrared colors for both models is similar under the 
assumption of an intermediate age burst as described above. With a total extinction of 
{\it A}$_{\rm V}$ = 16.2 magnitudes, the non-scattering model predicts ({\it B$-$I, 
I$-$H, H$-$K}\p) = (1.63, 1.96, 0.54), while the scattering model predicts ({\it 
B$-$I, I$-$H, H$-$K}\p) = (1.46, 1.60, 0.51).

A more important change, however, is that scattering reduces the effective optical depth 
by increasing the fraction of light originally emitted by the stars which actually 
emerges from the model.
This is particularly valuable because the mixed stars and dust model without scattering 
developed above cannot easily constrain the total bolometric luminosity of the stellar 
ensemble; it is relatively easy to hide luminosity at high optical depth where it cannot be 
seen in comparison to the stars at lower optical depth. One can therefore easily increase 
the ``geometric'' correction (the difference in magnitude between the true luminosity 
and that observed) at any given wavelength by simply specifying a sufficiently large 
total optical depth for the star/dust mixture. In the uniformly mixed stars and dust 
model without scattering and with a total optical depth of {\it A}$_{\rm V}$ = 50 mag, 
for example, the geometric correction to convert between the actual total stellar 
luminosity and that observed is nearly 4.8 magnitudes at B and 3.5 at K. For an ensemble 
with {\it A}$_{\rm V}$ = 16 magnitudes these numbers drop to 3.1 and 0.8, 
respectively.
For the ``dusty'' galaxy model with scattering, however, even if the total optical depth at 
B is as high as $\tau$=20 (corresponding to the {\it A}$_{\rm V}$ =16 magnitudes 
case described above), the geometric correction is only 2.4, and at K is 0.7. This effect 
is even more pronounced for other dust geometries; at {\it A}$_{\rm V}$ =16 magnitudes 
the geometric correction at B is 0.4 for stars with dust lying interior to the stars and 
1.8 for a ``starburst'' system with a centrally concentrated stellar distribution and dust 
mixed with the stars. Witt \& Gordon (1996) showed that this effect becomes even more 
pronounced when the scattering medium is assumed to be clumpy instead of homogeneous, 
in which case the effective optical depth may be decreased by additional factors of 2--4. 
The addition of the scattered light to the total observed light therefore makes the 
observed luminosity more indicative of the actual total luminosity.

Figures 6 and 7 show the three-dimensional space occupied by the three colors ({\it 
B$-$
I}), ({\it I$-$H}), and ({\it H$-$K\p}). The resulting cube has been rotated by 
roughly 
90$^{\circ}$ in order to illustrate the orthogonal nature of the extinction and dust 
emission curves. Note that stellar and QSO nuclear emission are well separated in the 
three-color representation. The ``dusty'' galaxy scattering model occupies nearly the 
same color space as the free-free emission model. The ({\it B$-$I}) colors of QSOs are 
similar to the expected 
colors of  moderate age (10--100 Myr) starburst clusters; the largest distinguishing 
characteristic between the two is that QSO nuclei are more luminous at long wavelengths. 
As a result, the primary advantage of the short wavelength ({\it B$-$I}) data is to allow 
the disentanglement of extinction effects. 

\begin{figure}[p]
\plotone{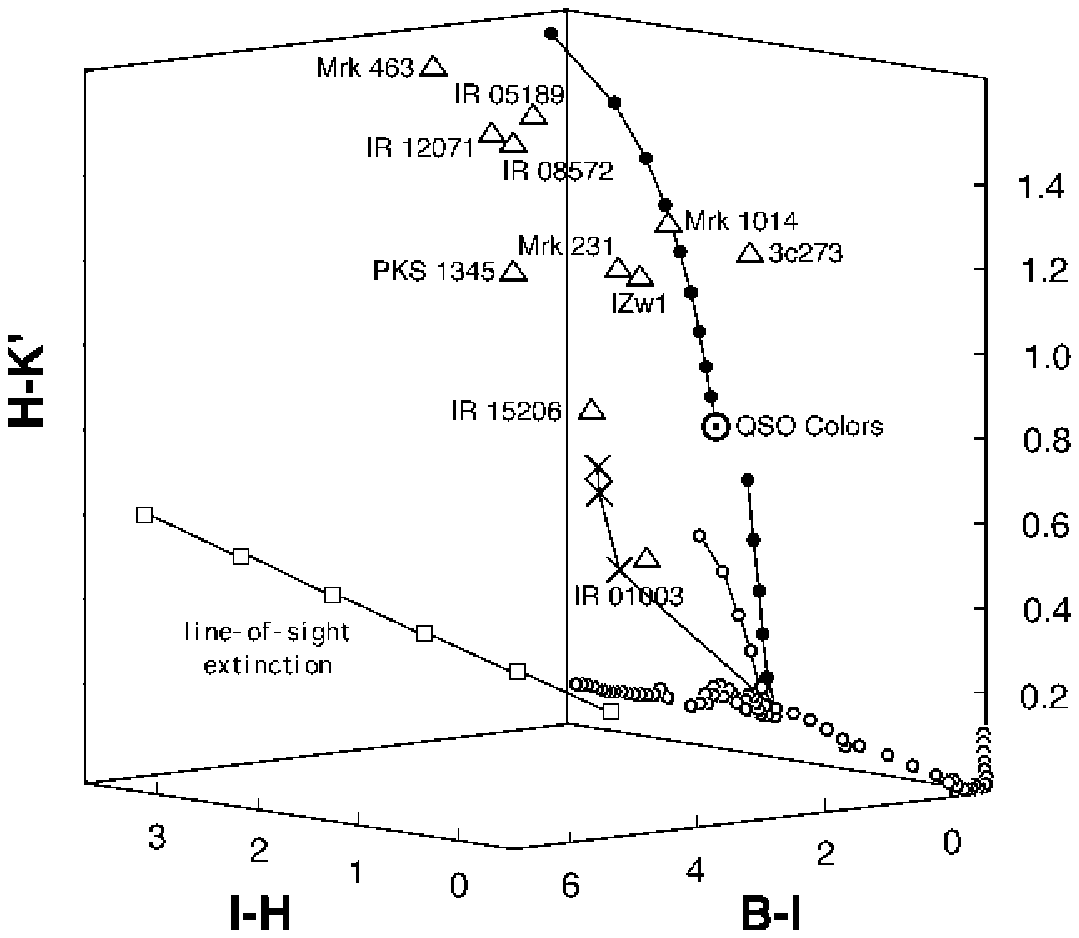}
\caption{({\it B$-$I}), ({\it I$-$H}), ({\it H$-$K\p}) color cube illustrating
the
colors of
the following emission sources: (1) optical QSO nuclear colors indicated by the
large,
dotted circle and (2) the evolutionary track of an instantaneous starburst aging
from 0
to 15 Gyrs, indicated by the open circles. It also shows the effects of the following
physical processes on the colors of stars and QSOs: (1) line-of-sight dust extinction
in
units of {\it A}$_{\rm V}$ = 1 magnitude (open squares), (2) free-free emission
with
an electron temperature of 20,000 K in increments of
20\% of the total flux at K\p (open joined circles), (3) emission from 800 K dust
in
increments of 10\% of the total flux at K\p (two sets of filled, joined circles
representing hot dust combined with either a QSO or young stellar SED), and (4)
emission from
uniformly mixed stars and dust, in units of {\it A}$_{\rm V}$=10, 30,
and 50 magnitudes ($\times$). Note that the line-of-sight dust extinction and thermal
dust reddening curves are nearly
orthogonal.}
\end{figure}

\begin{figure}[p]
\plotone{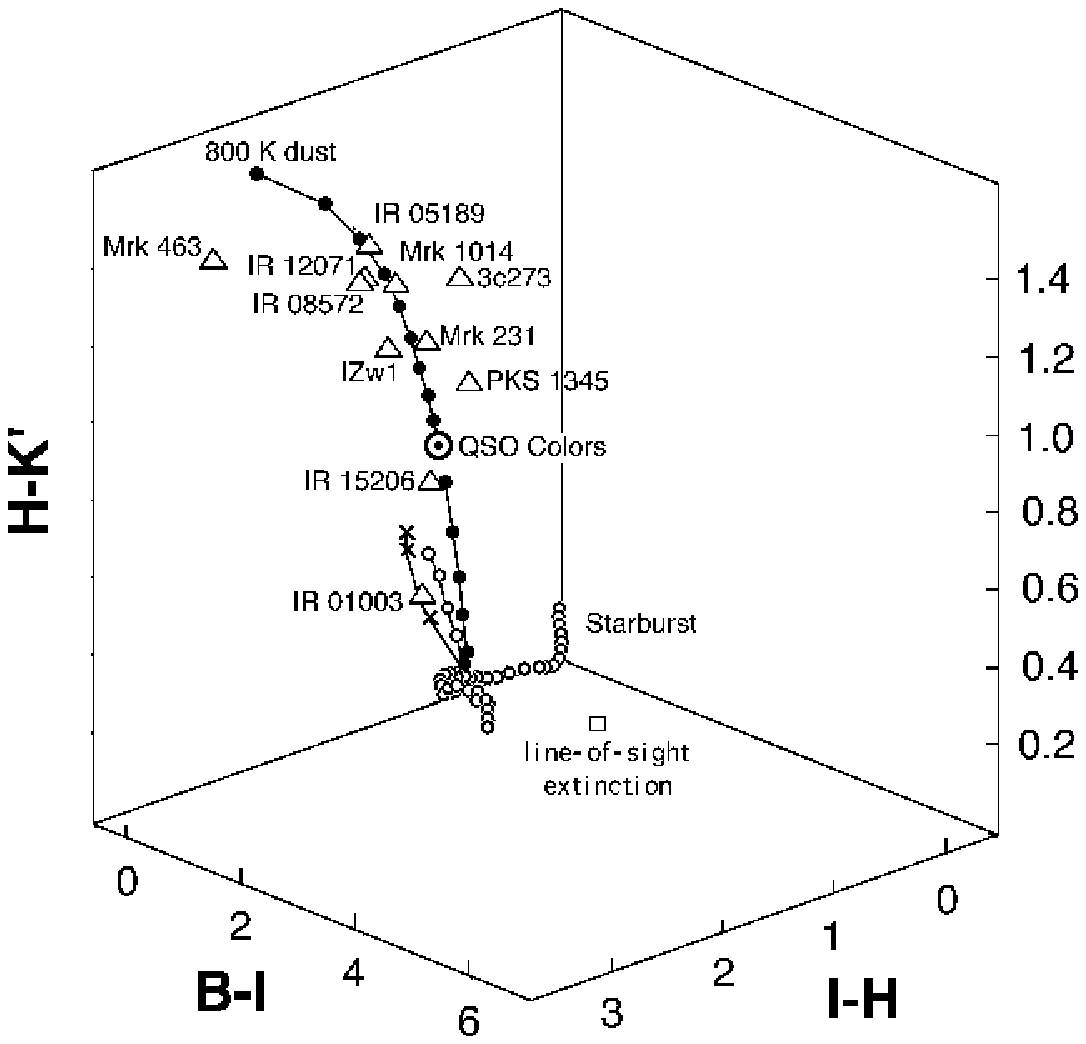}
\caption[Reddening-Orthogonal Optical/Near-Infrared Color Cube]{ Same as Figure 
6, but rotated such that the projected color basis is 
orthogonal to the line-of-sight dust extinction, i.e. line-of-sight extinction does not 
affect the position of points in this graph. Note the separation between QSO and stellar 
colors. Also, nearly all the warm ULIG nuclear colors are very similar to that of the 
infrared-loud QSOs, with the exception of IRAS{\ts}01003$-$2238.}
\end{figure}

The following discussion will concentrate primarily on Figure 7, which has been rotated 
such that the color basis projected onto the page is orthogonal to extinction (i.e. line-of-
sight extinction does not change the apparent position of the points). Since extinction is 
almost certainly present to a varying degree throughout these galaxies, it is important to 
eliminate as much as possible extinction effects. To distinguish between stellar emission 
and QSO activity, any processes capable of producing similar optical/near-infrared 
colors must be considered. In the following discussion, ``reddening'' will be used to mean 
processes which move points upward in the projected basis of Figure 7, primarily 
giving them larger values of ({\it H$-$K\p}), as opposed to the reddening caused by 
extinction. There are few processes which can make the QSO emission bluer. Large 
amounts of free-free emission (open circles, the last of which is pure free-free 
emission) can decrease ({\it H$-$K\p}) by only as much as 0.2. Likewise, free-free 
emission can redden the stellar colors only slightly. The most important effects are 
clearly that of hot dust and dust mixed with stars. The overall effect of mixed dust and 
stars is to 
redden the basic stellar colors. 
As discussed above, however, this reddening saturates when sufficiently high optical 
depths are reached. Furthermore, models of distributed stars and dust with both 
scattering and absorption indicate that it is difficult to redden starlight even with very 
high total optical depths (Witt et al. 1992,1996), and hence it is difficult to mimic the 
very red near-infrared QSO colors.


Since the QSO and stellar colors differ primarily at long wavelengths, the most serious 
problem occurs with reddening processes that affect only K\p and longer wavelength 
bands, such as thermal dust emission. If 800 K dust contributes as much as 55--60\% 
of 
the emission at K\p, then the resulting stellar/hot dust ensemble will have colors nearly 
identical to that of optically selected QSOs. Thankfully, some reprieve is noted in that the 
extremely red colors of hot dust can also redden the QSO nuclei, and hence any hot dust 
component around a QSO moves it even further away from the basic starburst colors. 
Unfortunately, the very red dust colors also imply that the stellar-hot dust emission 
will converge on the same colors as QSO-hot dust emission as the percentage of hot dust 
increases. For stellar colors, this effect is noticeable when hot dust accounts for $>$ 
70\% of the emission at K\p. Carico et al. (1990) found that for far-infrared bright 
and luminous ({\it f}$_{60} >$ 5 Jy; {\it L}$_{\rm ir} > 10^{11}$ {\it L}$_{\sun}$) 
galaxies which do not show signs of Seyfert activity, typically ({\it H-K\p})= 0.40 
(ranging from 0.16 to 0.57). Since these galaxies are believed to be powered 
predominantly by star formation, this represents the typical near-infrared colors 
observed for the most powerful of clearly identifiable starbursts.

Because the detailed SEDs of the high spatial frequency structure in the warm ULIGs are 
unknown, particularly at short rest wavelengths, it is not possible to apply K-
corrections to them. However, since the models of the emission processes, starbursts, 
and QSOs do have detailed SEDs, inverse K-corrections can be made to their rest-frame 
colors at the redshift of our targets. This is done by convolving the synthetic spectra 
with the known detector and filter bandpasses. The magnitude zeropoint calibration of the 
filters is derived using the Kurucz model spectrum of Vega (BC95). For brevity, 
Figures 6 and 7 are calibrated to the median ULIG redshift, z=0.12. The K-corrections 
affect the modeled stellar colors by $\delta$({\it B$-$I,I$-$H,H$-
$K\p})=(0.05,0.07,0.18) 
for a young (10 Myr) starburst, and by (0.40,0.11,$-$0.08) for an old (2 Gyr) 
population. The effects of K-corrections can therefore be quite large depending on the 
modeled population, and hence the representation of stellar colors can be considered to 
have a sizable uncertainty attached. Larger redshift values also shift the dust emission 
curves to a more vertical orientation, as the rest frame filter bandpasses become bluer.

\subsubsection{Nuclear Colors}

Also shown in Figures 6 and 7 are the colors of the ULIG nuclear regions (open 
triangles).
IRAS{\ts}01003$-$2238 seems to be unique in that it is most consistent with a young 
(10$^7$ yrs) stellar population mixed with local dust extinction with an optical depth 
of {\it A}$_{\rm V}$ magnitudes. It could also be a combination of hot young stars with 
20\% 
hot dust at K\p and {\it A}$_{\rm V}$ = 1 magnitude of extinction. 
IRAS{\ts}15206+3342 has an 
SED similar to that of a QSO reddened by {\it A}$_{\rm V}$ = 1 magnitude, but it is also 
similar to a 
young stellar population with a very large thermal dust component and {\it 
A}$_{\rm V}$ = 1--2 magnitudes of line-of-sight extinction. Given the large range of 
QSO colors it is not 
possible to differentiate between these two possibilities based on the optical/near-
infrared colors alone.

The remaining 9 galaxies all have SEDs characterized by large K\p excesses. The three 
optically identified QSOs (I{\ts}Zw{\ts}1, Mrk{\ts}1014, and 3c273) may be taken as 
representative of a subclass of ``infrared loud'' QSOs. Mrk{\ts}231, the most QSO-like 
of the ULIGs, has an SED nearly identical to that of the infrared-loud QSOs. Similarly, 
IRAS{\ts}{05189$-$2524}, IRAS{\ts}08572+3915w, and IRAS{\ts}12071$-$0444 
also 
have colors essentially identical to that of the infrared-loud QSOs seen through 
$A_{\rm V} \approx$  2 magnitudes of extinction. PKS 1345+12w has a smaller near-
infrared 
excess than the infrared-loud QSOs, but has the same colors as a typical QSO 
(characterized by our synthetic colors) seen through $A_{\rm V}$ = 2 magnitudes of 
extinction. Mrk 
463e has a larger excess at both H \& K\p than the others. This can be explained by 
emission from 800--900 K thermal dust which contributes 20--30\% more of the K-
band flux than exists in the infrared-loud QSOs.

Compared to the synthetic optically-selected QSO colors, both the ULIGs and the 
subsample of three infrared-loud QSOs have a noticeable K\p-band excess. This excess is 
2--3$\sigma$ greater than that expected considering the normal range of QSO colors 
(${\sigma}_{H-K{^{\prime}}}$=0.2; Elvis et al. 1994). However, it is likely that 
previous near-infrared and optical surveys have underestimated the ({\it H$-$K\p}) 
color of the QSO nuclei. Given that normal galaxy colors ({\it H$-$K\p}$<$0.2; 
Aaronson 
1977) are almost always bluer than QSO nuclei ({\it H$-$K\p}$>$0.8), studies that 
used 
large apertures to measure the QSO nuclear contribution are contaminated by underlying 
host galaxy light and therefore derive colors bluer than the actual nuclei. Also, several 
authors (Neugebauer et al. 1987, Sanders et al. 1989) have suggested that a hot dust 
component already contributes significantly to the near-infrared emission in optically 
selected QSOs. Given the presence of a strong far-infrared dust emission component and 
significant molecular gas content (see Sanders 1991), it would not be surprising if a 
larger reservoir of dust has also given rise to enhanced hot dust emission.

\subsubsection{Knot Colors}

In Table 3, we present photometry for the additional compact sources observed in the 
warm ULIGs. This includes both the galaxy ``nuclei'' which do not appear to have AGN-
like characteristics (e.g. IRAS{\ts}08572+3915e) and are likely to be the true stellar 
cores of galaxies, and the clumps of near-infrared emission which we associate with the 
compact star-forming ``knots'' observed in Paper I. The galaxy nuclei were measured 
using aperture photometry on the undeconvolved images, and have an uncertainty of 0.1 
magnitudes. The sky background was determined by using annular regions outside the 
aperture which appeared free of high frequency structure. The knot magnitudes were 
measured from the R-L deconvolved images in 
order to minimize confusion. Uncertainties in the photometry are introduced by the non-
linear, non-invertible nature of the deconvolution algorithm; these were checked by comparison with 
aperture photometry of the raw images. The knot magnitudes are uncertain by 0.15--
0.2 magnitudes.



\begin{deluxetable}{lrr}
\footnotesize
\tablenum{3}
\tablewidth{3.0truein}
\tablecaption{Additonal Structure}
\tablehead{
\colhead{Name} &
\colhead{\it m$_{\rm H}$} &
\colhead{\it m$_{\rm K^{\prime}}$}} 
\startdata
\cutinhead{IRAS 08572+3915}
15(east)$^{\dagger}$ & 16.2 & 16.0 \nl
\cutinhead{Mrk 231}
5/6 & 17.4 & 17.2 \nl
10/11/12 & 16.7 & 16.3 \nl
14 & 18.1 & 18.1 \nl
irA & 17.8 & 17.6 \nl 
\cutinhead{Pks 1345+12}
6(east)$^{\dagger}$ & 15.8 & 15.5 \nl
\cutinhead{Mrk 463}
19 & 17.3 & 16.8 \nl
20(west)$^{\dagger}$ & 14.3 & 13.8 \nl
21 & 16.9 & 16.7 \nl
28/29/30 & 18.0 & 17.6 \nl
\cutinhead{IRAS 15206+3342}
1 & 17.6 & 17.4 \nl
2 & 18.0 & 17.0 \nl
3/4/5 & 17.6 & 17.1 \nl
7/8 & 18.1 & 17.4 \nl
9$^{\dagger}$ & 16.9 & 16.2 \nl 
10 & 18.8 & 18.1 \nl
11 & 17.9 & 17.3 \nl
12/13 & 17.5 & 17.2  \nl
14 & 19.1 & 18.7 \nl
15/16 & $<$20.6 & $<$19.8 \nl
17 & 18.1 & 17.5 \nl
18 & $<$20.4 & $<$19.2 \nl
19 & 18.2 & 18.0 \nl
\tablecomments{$^{\dagger}$ denotes probable second (non-active) nucleus in double-nuclei systems. Names taken from Surace et al. (1998) and should be referred to in the form (galaxy name):SSVVM(number).
Names with multiple numbers indicate the given magnitude is an integrated value for the named knots.}
\enddata
\end{deluxetable}

The colors of the starburst ``knots'' found in Paper I were examined and initially 
revealed anomalous colors: while the knots had ({\it B$-$I}) and ({\it H$-$K\p}) 
colors that 
were consistent with stars, the ({\it I$-$H}) colors were far larger (typically ({\it 
I$-$
H})$>$ 2.5) than could be explained by any plausible combination of stellar colors and 
reddening effects. A careful examination of the data revealed that this was most likely due 
to aperture effects. The resolution of the optical data is typically 5-10 times higher 
than that of the infrared data. As a result, the apertures used to measure the optical data 
are much smaller. It is likely that the larger apertures used for the near-infrared data 
also include some additional underlying emission that may not be related to the optical 
knots, hence the apparent increase in luminosity between the optical and near-infrared 
data. However, as noted previously, the largest differences in emission and obscuration 
mechanisms lie in the optical ({\it B$-$I}) and near-infrared ({\it H$-$K\p}) colors. 
The 
purely optical and purely infrared knot colors should be fairly certain since data of 
different resolutions are not mixed. Therefore, the knot colors in the 2-color ({\it B$-
$
I}) vs. ({\it H$-$K\p}) plane are presented in Figure 8. 

\begin{figure}[p]
\plotone{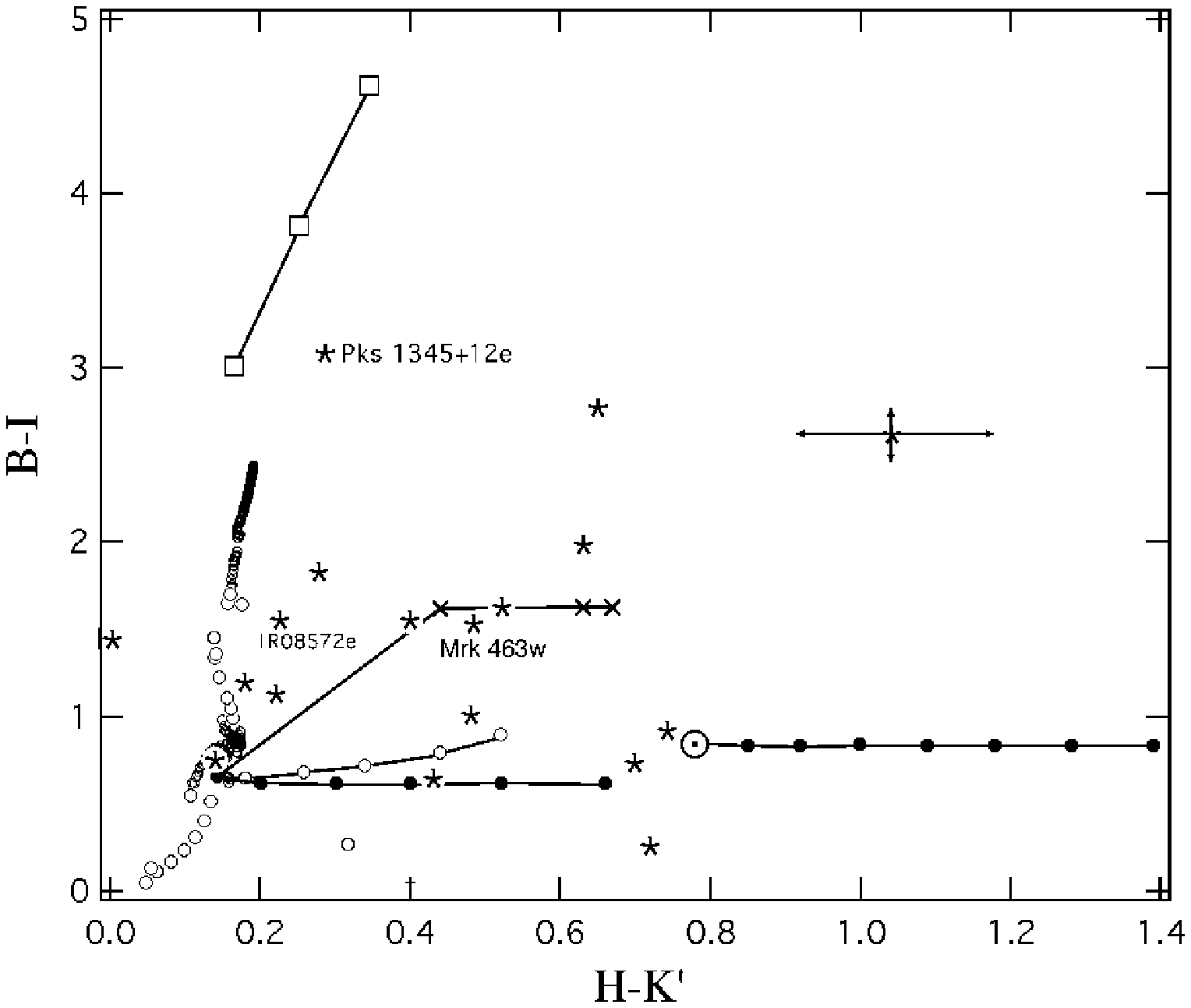}
\caption[Near-Infrared Knot Colors]{ The ({\it B$-$I}),({\it H$-$K\p}) color 
plane, with the same reddening 
mechanisms, stellar and QSO colors as in Figure 6. The star-forming knots are 
marked 
with stars and the stellar galaxy nuclei are labeled.}
\end{figure}

The colors of the galaxy nuclei appear very similar to that of stars with mild reddening. 
Pks{\ts}1345+12e has colors similar to that of an old stellar population with {\it 
A}$_{\rm V}$=1--2 magnitudes. IRAS{\ts}08572+3915e and Mrk 463w appear 
similar to a 
somewhat younger stellar population (1--2 Gyrs), but have a slight near-infrared 
excess. In the case of Mrk 463w this may not be surprising as this galaxy core shows 
both Seyfert 2 activity and numerous knots of star formation near it, indicating that 
there may be considerable new star formation in it, resulting in a thermal dust 
contribution. There is a surprisingly wide range in ({\it H$-$K\p}) colors for the 
putative starburst knots themselves. For comparison, we consider larger samples of 
luminous infrared galaxies (LIGs; $10^{11} < L_{\rm ir} < 10^{12}$ {\it 
L}$_{\sun}$; Carico 1988, 1990). These are the most far-infrared luminous galaxies 
in which it is fairly certain that star formation is the primary contributor to the far-
infrared luminosity, and as such represent the most extreme known starburst galaxies. 
The distribution in knot colors is very similar to that of the LIG nuclear regions, ({\it 
H$-$K\p})= 0.25--0.6, not counting the few LIGs known to be Seyferts and which are 
therefore likely to have a strong non-stellar component. 

Carico et al. (1990) found that the ``disk'' ({\it H$-$K\p}) colors of luminous IRAS 
galaxies were typically ({\it H$-$K\p})=0.25 (ranging from 0.08 to 0.37). It is 
unclear 
what stellar components were integrated into this ``disk'' light --- clearly there must 
be some contamination by star-forming regions, which probably accounts for the 
difference between these colors and the value of {\it H$-$K\p}=0.16 found by Aaronson 
(1977). 
The colors of the underlying host galaxies of the warm ULIGs were examined by 
subtracting the contribution of the putative nucleus as well as any additional 
contribution from identified point-like sources from the integrated galaxy flux. In this 
case the contribution from the putative nucleus was taken as the flux inside the central 
2.5 kpc. This should ensure that all the nuclear contamination of the underlying galaxy 
colors is removed and, although this process also eliminates some of the host galaxy's 
flux, should leave the colors unchanged. Most of the colors so derived for the warm ULIGs 
appear similar to those found by Carico et al. (1990), ranging from {\it H$-
$K\p}=0.09 
in IRAS{\ts}05189$-$2524 to nearly {\it H$-$K\p}=0.7 in IRAS{\ts}08572+3915 
(the 
latter probably being strongly affected by dust) with a median of {\it H$-$K\p}=0.4. 
However, a few of the systems seem to show peculiarly red colors, namely the optically 
selected QSOs. This is probably the result of residual nuclear contamination caused by 
the extended wings of the bright quasar nuclei.

\subsection{Contribution of putative ``nuclei'' to the bolometric luminosity}

ULIGs are defined by their extremely high infrared luminosities ({\it L}$_{\rm 
ir}>$10$^{12}${\it L}$_{\sun}$, equivalent to the bolometric luminosities of 
optically selected QSOs). If the source of the majority of this luminosity were an AGN, 
then it is likely that after expulsion of the enshrouding dust the AGN would appear 
similar to an optical QSO. Although the broad-band optical/near-infrared colors of the 
warm ULIG nuclei (as defined in \S 3.1, the dominant compact emission sources at K\p \ 
spatially 
coincident with previously identified optical nuclei) appear very similar to that of QSOs, 
many of the nuclei are less luminous at these wavelengths than QSOs. Three of our 12 
ULIGs (the optically identified QSOs I{\ts}Zw{\ts}1, Mrk{\ts}1014, and 3c273) 
already have a substantial fraction of their {\it L}$_{\rm bol}$ in a QSO-like SED at 
short wavelengths (i.e. ``the blue bump''). The nucleus of the broad-line object 
IRAS{\ts}07598+6508 also meets the minimum {\it B}-band luminosity criterion for 
QSOs (Surace, unpublished).
Dereddening of the observed optical emission by the optically estimated extinction 
(typically {\it A}$_{\rm V}$ = 2 magnitudes, Surace et al. 1998) is insufficient to 
raise the 
estimated optical luminosity of the remaining eight AGN to the canonical {\it M}$_{\rm 
B}$ = $-$22.1 luminosity criterion for optical QSOs except in the case of the nucleus of 
Mrk 231 (dereddened {\it M}$_{\rm B}$ = $-$23 ). Dereddened B-band luminosities 
of 
the remaining seven range from {\it M}$_{\rm B}$=$-$15.5 for 
IRAS{\ts}08572+3915 
to {\it M}$_{\rm B}$ = $-$22 for IRAS{\ts}15206+3342. The median value is {\it 
M}$_{\rm B}$ = $-$20.2, which is typical of Seyferts (Osterbrock 1989). 

Due to the much lower extinction in the near-infrared, it is possible that the nuclei in 
many cases are seen directly in the near-infrared (which is supported by their 
essentially unresolved near-infrared morphology) and hence the observed K\p-band 
luminosity may be more indicative of their true luminosity, although it is also sensitive 
to hot dust emission, as in \S 3.2. Using the observed colors for I{\ts}Zw{\ts}1, 
Mrk{\ts}1014, and 3c273 (the infrared-loud optically-selected QSOs), {\it M}$_{\rm 
B}$=$-$22.1 is equivalent to a QSO K\p magnitude of {\it M}$_{\rm K^{\prime}}$=$-
$
25.8; using the synthetic optical QSO colors it is {\it M}$_{\rm K^{\prime}}$=$-
$25.2 
at the median redshift {\it z} = 0.12. IRAS{\ts}05189$-$2524 and Mrk 463e (which 
are 
also broad-line objects) now qualify as QSOs based on their K\p luminosity. The 
remaining five of the warm ULIGs still fall short of QSO near-infrared luminosities; 
their infrared luminosities range from {\it M}$_{\rm K^{\prime}}$=$-$23 (IRAS 
01003-2238) to {\it M}$_{\rm K^{\prime}}$=$-$24.7 (IRAS{\ts}12071-0444), 
with a mean of {\it M}$_{\rm K^{\prime}}$=$-$23.8. Therefore, while in 58\% 
(7/12) 
of the warm ULIGs the dereddened near-infrared nuclear luminosity is similar to an 
optically selected QSO and is probably the source of the majority of the bolometric 
luminosity, if we assume only a simple foreground reddening screen then the remaining 
five warm ULIG nuclei appear similar to lower luminosity Seyferts and fall an order of 
magnitude short of QSO luminosities. It is notable, however, that in these underluminous 
systems both the nuclei and the star-forming knots are insufficiently luminous at 
optical/near-infrared wavelengths (even after dereddening) to account for the high 
bolometric luminosity. Either a large fraction of the total luminosity must arise in a 
source or sources so heavily extinguished as to escape detection at {\it both} optical and 
near-infrared wavelengths, or alternatively the ``knots''/nuclei we have already 
detected must have intrinsic SEDs with an unusually strong far-infrared component in 
comparison to the putative nuclei in the seven ULIGs which have dereddened 
optical/near-infrared luminosities consistent with QSOs.

Until now we have assumed that both the optical and near-infrared emission reached the 
observer through a common path. The apparent optical and near-infrared 
underluminosity in five of the warm ULIGs relative to QSOs may be understood as a 
putative AGN being seen through a combination of patchy extinction and scattering. The 
nucleus might in some cases be seen directly in the near-infrared due to the lower near-
infrared extinction, but primarily through scattered light in the optical (which may 
then be additionally reddened by extinction along the scattered path). In more extreme 
cases the line-of-sight column density may be so high that the AGN is seen only in 
scattered light even at near-infrared wavelengths. Since the extinction estimates are 
most heavily influenced by the short wavelength colors (where extinction effects are 
greatest) the extinction estimates would be dominated by that along the low-extinction 
scattered paths, leading to estimated extinctions that are much too low. Attempts to 
derive true luminosities based on the optical light would then be futile if they are 
dominated by scattered light which is dependent on the scattering geometry. 

Polarization studies (Young et al. 1996) which included 6 of our warm ULIGs have 
actually found evidence for patchy obscuration in the vicinity of the scattering region. In 
particular, their conclusions for IRAS{\ts}05189$-$2524 agree with the conclusion 
presented in \S 3.1 based on morphology that the emitting region is seen directly in the 
near-infrared, but with an optical scattered component with mild optical extinction and 
a sizable contribution from starlight, which is probably a result of the circumnuclear 
starburst surrounding the AGN (Paper I). A similar result is found for Mrk 463e, 
whose optical/near-infrared  morphology shows that the optical emission is dominated 
by a spatially distinct region to the north of the near-infrared nucleus. Recent HST 
results indicate that this emission is that of a highly polarized Seyfert 1 ``hidden'' 
nucleus (Tremonti et al. 1997). This  agrees with the polarized SED found by Young et 
al. (1996) who concluded that the line-of-sight extinction to this nucleus was A$_{\rm 
V}$=17 magnitudes; this high extinction agrees with our observed morphology. The 
steeply rising 
polarization in the near-infrared of IRAS{\ts}08572+3915w (Young et al. 1996) may 
indicate that it is so deeply buried that almost none of the optical emission arises from 
the AGN, and that even at near-infrared wavelengths the emission is predominantly 
scattered and not line-of-sight. 
The case of IRAS{\ts}08572+3915w reinforces the point that morphologically the 
galaxies with the least luminous nuclei also tend to have more complex optical 
morphology, suggestive of patchy scattering. Additionally, recent high spatial resolution 
observations of nearby Seyfert galaxies have indicated the importance of patchy 
extinction in differentiating Seyfert classes (Malkan et al. 1998). Given that the far-
infrared emission in the 
warm ULIGs is 50--90\% of the bolometric luminosity, then this implies that the 
covering factor of a putative QSO must be of the same order. Patchy extinction must 
therefore be present; only in those cases where the nuclei are seen along a direct line of 
sight would the unreddened luminosities be consistent with QSOs.

In terms of the previous color analysis which assumed a simple single-path model for 
the optical/near-infrared emission, correcting for the effects of patchy extinction and 
scattering would be to drive the nuclear colors even further from those of stars, thus 
further eliminating some of the degeneracy in the color diagram. The apparent 
enhancement at shorter wavelengths due to the addition of scattered light would result in 
the derivation of colors that are too blue --- correction for this further reddens the 
colors. While most polarization studies detect only small fractional polarizations ($<$ 
5\%), they also use very large apertures (5--8\arcsec), which encompass large 
fractions of more distant galaxies like the warm ULIGs studied here (Young et al. 1996). 
The beam dilution can be estimated by comparing the B-band emission of the putative 
nuclei in the HST/WFPC data to the total flux encompassed in a 5\arcsec \ aperture.
If the polarized light is actually associated with the high surface brightness features 
observed with HST then the beam dilution correction may be anywhere from 5$\times$ 
in Mrk 463e to as high as 50$\times$ in IRAS 08572+3915, with 10--15$\times$ 
being more typical. In this case, a significant fraction of the observed optical nuclear 
emission may be scattered, affecting the analysis as mentioned previously. 

\subsection{Contribution of putative star-forming ``knots'' to the bolometric 
luminosity}

In some of the objects in our sample of warm ULIGs the starburst knots (compact 
emission regions not identified as ``nuclei'') may contribute as much as 
25\% of the bolometric luminosity.
Since the ULIG knots appear to be powerful bursts of star-formation similar to the 
circumnuclear starbursts found in the LIGs, they may have similar SEDs.
The relationship between K\p and far-infrared luminosity in the LIGs can then be used to 
estimate a far-infrared luminosity of the ULIG knots based on their observed near-
infrared luminosity. From Carico et al. (1990) the empirical relation 

\begin{equation}
{\rm log} L_{\rm ir}=-{{M_{\rm K^{\prime}}-6.45}\over{2.63}}
\end{equation}

\noindent is derived for the subset of LIGs without any evidence for a strong AGN 
component; this relates the ``nuclear starburst'' near-infrared luminosity to {\it 
L}$_{\rm ir}$. For those warm ULIGs where it is possible to observe bright near-
infrared knots, the extrapolated total far-infrared knot luminosity ranges from {\it 
L}$_{\rm ir}$ = 10$^{10.27}$ {\it L}$_{\sun}$ in Mrk 231 to as high as log {\it 
L}$_{\rm ir}$ = 10$^{11.6}$ {\it L}$_{\sun}$ in IRAS{\ts}15206+3342. Previous 
estimates of the knot bolometric luminosities (e.g. Surace et al. 1998) were 
appreciably lower (0.1--2.5 $\times$ 10$^{10}$ {\it L}$_{\sun}$) since they were 
based on their {\it B} magnitudes and a simple foreground reddening model applied to a 
purely stellar (no dust) starburst. The new near-infrared observations reveal a near-
infrared excess in the knots similar to other infrared-luminous starbursts which 
results 
in higher estimated bolometric luminosities. Nevertheless, the net result is that the knot 
contribution to {\it L}$_{\rm bol}$ is much less than that of the putative nuclei, which 
are thus likely to be the dominant energy source in the warm ULIGs.

\section{Conclusions}

We have presented high spatial resolution images of a complete sample of ``warm'' 
ultraluminous galaxies. From these images we conclude the following:

1) The warm ULIGs have a near-infrared structure very similar to that seen optically, 
with tidal features and some clumpy emission coincident with the star-forming knots 
seen optically. Most of these knots have low line-of sight extinctions. We have failed to 
detect significant numbers of new near-infrared knots indicating that any additional 
knots are either extinguished by greater than {\it A}$_{\rm V}$=25 magnitudes, are 
intrinsically 
less luminous, or are distributed differently from the optical knots.

2) The starburst knots have a wide range of ({\it H$-$K\p}) infrared excesses. These 
excesses span the range seen in the less luminous LIGs. Adopting the SED of the LIGs as 
representative of the most luminous star-forming regions, then the knots seen in the 
warm ULIGs have total, estimated far-infrared luminosities in the range of 10$^9$--
10$^{11.5}$ {\it L}$_{\sun}$. This is similar to the range found in non-active 
infrared galaxies. 

3) A single ``knot'' of emission, spatially coincident with an optically identified 
``nucleus'', increasingly dominates the emission at long wavelengths. This putative 
nucleus is generally unresolved at K\p.

4) The colors of these putative nuclei are nearly identical to that of optically selected 
QSOs with an added hot ($\approx$800 K) dust component contributing 10--30\% of 
the K\p emission seen through {\it A}$_{\rm V}$=1--2 magnitudes of line-of-sight 
extinction. These same 
colors are identical to ``infrared loud'' quasars. They could, however, also be produced 
by a powerful starburst 10--100 Myrs in age with 70\% or more of the flux at K\p 
originating in hot dust. This would be a greater percentage than that observed in other, 
less luminous starburst galaxies.

5) Seven of the sample of 12 warm ULIGs have dereddened ``nuclear'' optical and near-
infrared luminosities consistent with that of optically-selected QSOs. In the remaining 
five cases the dereddened luminosities of the putative nuclei are similar to that of 
Seyferts. If the underlying emission source is a QSO, then this could be explained by the 
optical emission being seen primarily through scattered light along less extinguished 
lines of sight. 

6) The powerful Wolf-Rayet galaxy IRAS{\ts}01003$-$2238 appears unique among 
the 
warm ULIGs: it is pointlike, has no obvious tidal features or extended host, and is under-
luminous at both optical/near-infrared wavelengths. 

7) The dereddened ``nuclear'' luminosities of the sample of warm ULIGs are sufficiently 
high that in 58\% (7/12) of the cases they could provide most of the bolometric 
luminosity. In all cases except IRAS{\ts}15206+3342 their contribution to the 
bolometric luminosity is likely to greatly exceed that of the more widespread star-
formation detected so far. 

\acknowledgements

We would like to thank Kevin Jim, Mark Metzger, Joe Hora and Klaus Hodapp for their 
advice and work on the fast tip/tilt secondary and the QUIRC infrared camera. We would 
also like to thank Buzz Graves and Malcolm Northcott for their knowledge on the 
characteristics of AO systems. We also thank Lee Armus and B.T. Soifer for helping with 
the early groundwork for this project. Finally, we thank Bill Vacca, Alan Stockton, Steve 
Willner, and an anonymous referee whose useful comments helped strengthen both the 
material and presentation of this paper.
J.A.S. and D.B.S. were supported in part by NASA grant NAG5-3370.

\clearpage

\appendix
\section{Notes on Individual Objects}

{\it I{\ts}Zw{\ts}1} --- a Palomar-Green (PG) QSO. Both spiral arms are clearly 
detected 
in the near-infrared. The QSO nucleus is situated on an extremely bright elongated 
stellar galaxy core.

{\it IRAS{\ts}01003$-$2238} --- A powerful Wolf-Rayet galaxy (Armus et al. 
1988) 
with several small knots of star-formation seen optically. The near-infrared data 
reveals no features at all.

{\it Mrk{\ts}1014} --- a PG QSO observed optically to have several knots and linear 
features very close to the nucleus (Paper I), and a very large tidal arm (MacKenty \& 
Stockton 1984). The single-armed tidal 
feature to the east of the optical nucleus is clearly detected in the near-infrared.

{\it IRAS{\ts}05189$-$2524} --- the extended structure of the host galaxy is clearly 
detected at H. Under careful scrutiny, the eastern and southern loops can also be detected. 
Young et al. (1996) report broad polarized H$\alpha$ and polarization evidence that the 
nucleus is seen directly in the near-infrared, confirmed by our detection of an 
apparently unresolved nucleus. This contrasts with the optical morphology presented in 
Paper I which showed a double peaked feature that appeared most consistent with a single 
object bisected by a dust lane.

{\it IRAS 07598+6508} --- our data for this object is poor. However, we do not 
readily 
detect the tidal structure and star-forming knots seenoptically to the south and east of 
the 
nucleus (Boyce et al. 1996), probably indicating very blue colors.

{\it IRAS{\ts}08572+3915} --- in the near-infrared, both nuclei can be clearly seen, 
as 
can the two tidal tails extending to the north and east. The western nucleus appears 
similar to a point superimposed on an extremely bright, extended background. The 
eastern nucleus is oval in shape and is very similar both in infrared and optical 
appearance.

{\it Mrk{\ts}231} --- this system is dominated in the near-infrared by the active 
nucleus seen optically. Additionally, the optical knots in the southern ``horseshoe'' are 
detected, and are even more distinct in the deconvolved images. Additionally, we seem to 
have detected at least one near-infrared knot not seen optically ``irA''.

{\it Pks{\ts}1345+12} --- the extended galaxy is clearly seen and is similar in 
appearance to previous optical images, and has no fine structure. The underlying host 
galaxy is very luminous and thus the nuclei contribute less than half to the total 
near-infrared luminosity.

{\it Mrk{\ts}463} --- this system is dominated at K\p by the eastern nucleus. The peak 
of 
the emission is located at knot 15 of Surace et al. (1998). This seems to imply that this 
knot is the true nucleus, and that the optically bright spot to the north is an excitation 
feature. This is probably the source of the polarized broad lines reported by Tremonti et 
al. (1997). The extended tidal remnants are detected, as are the star forming knots in 
these tails (primarily to the north and east).

{\it IRAS{\ts}15206+3342} --- as noted in Surace et al. (1998), this system 
possesses 
4 extremely bright optical knots, as well as more than a dozen less luminous knots. In 
the deconvolved near-infrared images, these knots are readily visible. The northeastern 
knot seems to dominate at K\p. No additional knots are seen in the near-infrared.

\newpage

\end{document}